\documentclass[%
 aip,
 amsmath,amssymb,
 reprint,%
]{revtex4-1}

\usepackage{graphicx}
\usepackage{dcolumn}
\usepackage{bm}
\usepackage[utf8]{inputenc}
\usepackage[T1]{fontenc}
\usepackage{mathptmx}
\usepackage{etoolbox}
\usepackage{enumitem}
\usepackage{ulem}

\usepackage[dvipsnames,svgnames,table]{xcolor}

\makeatletter
\def\@email#1#2{%
 \endgroup
 \patchcmd{\titleblock@produce}
  {\frontmatter@RRAPformat}
  {\frontmatter@RRAPformat{\produce@RRAP{*#1\href{mailto:#2}{#2}}}\frontmatter@RRAPformat}
  {}{}
}%

\makeatother
\begin{document}

\preprint{AIP/123-QED}

\title[A dynamical model for Brownian molecular motors driven by inelastic electron tunneling]{A dynamical model for Brownian molecular motors driven by inelastic electron tunneling}
\author{Federico D. Ribetto}
\affiliation{Instituto de F\'isica Enrique Gaviola (CONICET) and FaMAF, Universidad Nacional de C\'ordoba, Argentina}
\affiliation{Departamento de F\'isica, Universidad Nacional de R\'io Cuarto, Argentina}
\author{Sebasti\'an E. Deghi}
\affiliation{Instituto de F\'isica Enrique Gaviola (CONICET) and FaMAF, Universidad Nacional de C\'ordoba, Argentina}
\author{Hern\'an L. Calvo}
\affiliation{Instituto de F\'isica Enrique Gaviola (CONICET) and FaMAF, Universidad Nacional de C\'ordoba, Argentina}
\author{Ra\'ul A. Bustos-Mar\'un}\email{rbustos@famaf.unc.edu.ar}
\affiliation{Instituto de F\'isica Enrique Gaviola (CONICET) and FaMAF, Universidad Nacional de C\'ordoba, Argentina}
\affiliation{Facultad de Ciencias Qu\'imicas, Universidad Nacional de C\'ordoba, Argentina}

\date{\today}

\begin{abstract}
In recent years, several artificial molecular motors driven and controlled by electric currents have been proposed. Similar to Brownian machines, these systems work by turning random inelastic tunneling events into a directional rotation of the molecule. Despite their importance as the ultimate component of future molecular machines, their modeling has not been sufficiently studied. Here, we develop a dynamical model to describe these systems. We illustrate the validity and usefulness of our model by applying it to a well-known molecular motor, showing that the obtained results are consistent with the available experimental data. Moreover, we demonstrate how to use our model to extract some difficult-to-access microscopic parameters. Finally, we include an analysis of the expected effects of current-induced forces (CIFs). Our analysis suggests that, although nonconservative contributions of the CIFs can be important in some scenarios, they do not seem important in the analyzed case. Despite this, the conservative contributions of CIFs could be strong enough to significantly alter the system's dynamics.
\end{abstract}

\maketitle

\section{\label{sec:intro}Introduction}

Any molecular system capable of rotating against a surface or solid can be considered a molecular rotor~\cite{kottas2005}.
However, for this type of system to be considered a rotational (or rotary) motor, it has to be capable 
of producing useful work. A key element to fulfill this condition is to have the means to drive the molecular rotor
unidirectionally in a controlled way~\cite{joachim2015, tierney2011a}, which requires an external power source according to the laws of thermodynamics.

Various energy sources, including chemical~\cite{kinosita2000, leigh2003, bauer2008, akimov2013}, light~\cite{koumura1999, vandelden2005, schonborn2009, kistemaker2015, wilcken2018, ikeda2019, romeo-gella2021}, and electrical~\cite{tierney2011a, kudernac2011, perera2012, echeverria2014, mishra2015, eisenhut2018, kugel2018, simpson2019, stolz2019, zhang2019, ren2020, wu2020}, have been theoretically and experimentally studied. However, compatibility with current microelectronic technology makes electric power sources particularly appealing for controlling these devices.

\begin{figure}[h]
  \centering
  \includegraphics[width=\columnwidth]{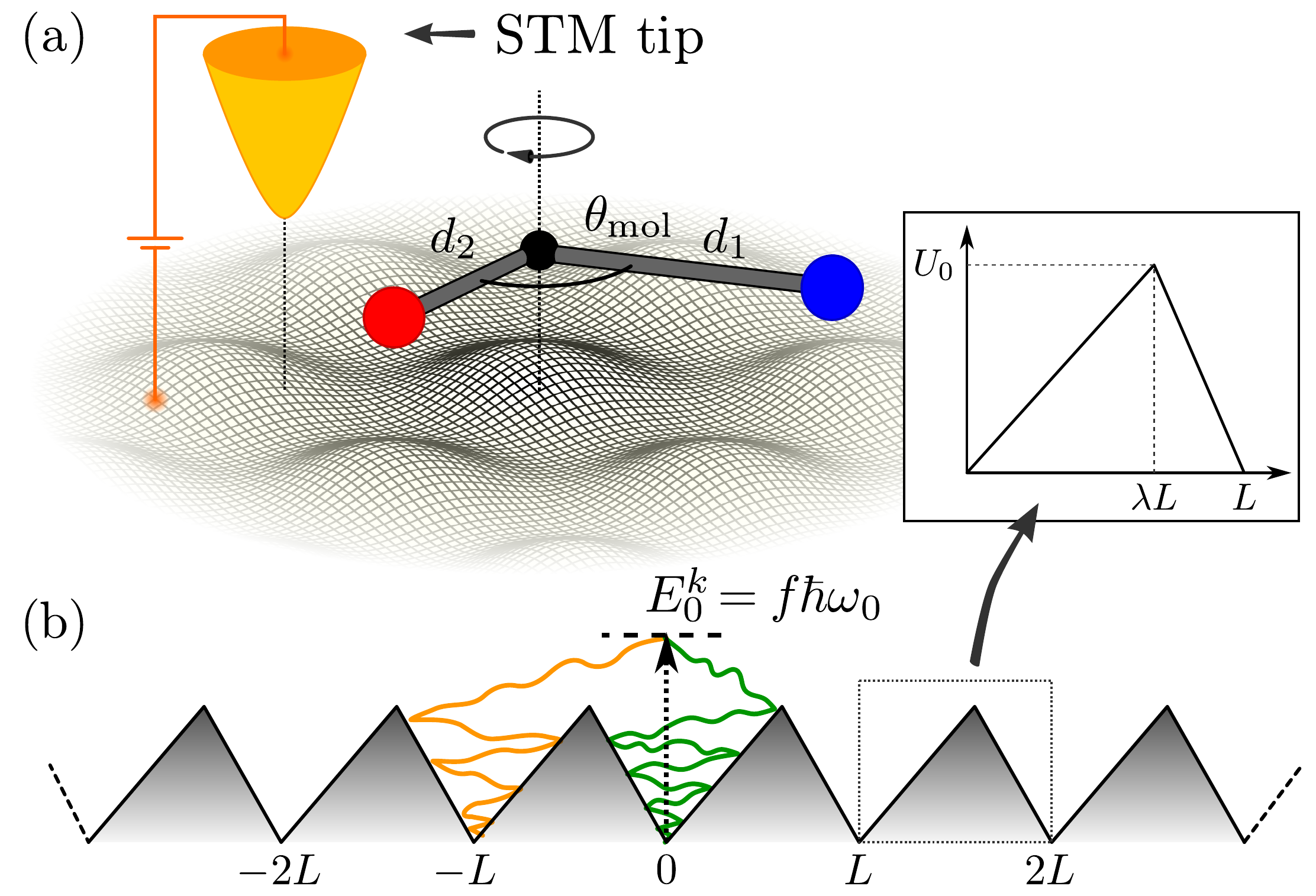}
  \caption{\label{fig:scheme}
 (a) Scheme of a typical Brownian molecular motor driven by inelastic electron tunneling (IET) events. A molecule, adsorbed on a metallic surface, is able to rotate when an electric current flows between the tip of a STM and the substrate.
  (b) Asymmetric potential landscape associated with the molecule's rotation. Each minimum corresponds to one of the equilibrium positions of the molecule, determined by its interaction with the metallic surface.
  In our model, the initial rotational kinetic energy $E^k_0$ is set equal to a fraction $f$ of the total energy transferred from the IET event, $\hbar \omega_0$.
  Green and orange lines show two molecular trajectories, each with one of the two possible directions of the initial rotational velocity.}
\end{figure}

The first experimental demonstration of an electrical (and rotational) molecular motor was achieved by Tierney and
coworkers, as reported in Ref.~\onlinecite{tierney2011a}.
In this experiment, the electrically driven rotation of a simple asymmetric molecule adsorbed on a surface was directionally biased.
More specifically, the authors managed to rotate individual molecules of the thioether butyl methyl sulfide (BuSMe) adsorbed on a Cu(111) surface by using the electron current from a scanning tunneling microscopy (STM) tip which
played the role of an electrode [see Fig.~\ref{fig:scheme}(a) for a simplified scheme of the experimental setup for this kind of molecular motors].
These molecules can hop in either the clockwise (CW) or counterclockwise (CCW) direction between six equidistant equilibrium positions set by the substrate's symmetry. They become chiral when adsorbed because of its asymmetric arms, and exist in two different forms
(enantiomers)~\cite{tierney2011a, murphy2014}.
Such a property is reflected in an asymmetric rotational potential energy surface, like the one shown in Fig.~\ref{fig:scheme}(b), as demonstrated by DFT calculations 
performed by the same group~\cite{tierney2011b}. 

The molecule's rotation has been attributed to the inelastic electron tunneling (IET) excitation of a C-H stretching mode and subsequent intramolecular vibrational energy redistribution~\cite{tierney2009, murphy2014}.
These occasional events excite the motor into a state of high energy compared to the equilibrium
torsional potential amplitude, which initially leads to rotations in both directions.
However, due to the torsional potential's asymmetry, there is a preference for one of the directions as the motor relaxes into one of its equilibrium positions. This preference is quantified by means of 
the system's \textit{directionality}, which compares the number of CW and CCW rotations (we
will further discuss on its definition in Sec.~\ref{sec:dir}).
This quantity is related to the \textit{distribution of hop angles} of the molecule, i.e., the frequency count of hopping events among the different minimum energy configurations. Both of these quantities were analyzed
and their values reported in Ref.~\onlinecite{tierney2011a}.

With this picture in mind, the measured directionality reflects the asymmetry
of the static torsional potential energy landscape~\cite{tierney2011a, murphy2014}, which can be modeled as a
sawtooth-shaped function, as shown in Fig.~\ref{fig:scheme}(b). As suggested by the authors, this phenomenon can be thought of as equivalent to a Brownian motor driven by temperature oscillations (also known as temperature ratchet~\cite{reimann2002, kay2007, hanggi2009}), where the sporadic electrical excitation of a C-H stretching mode is equivalent to the system's periodic increase in temperature.
The theory of Brownian ratchets states that these systems require both an asymmetric rotational potential and a
thermodynamical nonequilibrium source (see Ref.~\onlinecite{ren2020} and references  therein), conditions that are fulfilled in this kind of molecular motors.

A similar experiment using a simpler molecule was carried out by Stolz \textit{et al.}~\cite{stolz2019}, who developed an electrically driven motor composed of an acetylene molecule (C$_2$H$_2$) anchored to a chiral surface of PdGa(111). More complex molecules have also been used. For instance, Perera \textit{et al.} developed a molecular motor adsorbed on a Au(111) surface consisting of a five-arm rotor and a tripodal stator, both connected by a ruthenium atom~\cite{perera2012}.
For further examples on electrically-driven molecular motors see Refs.~\onlinecite{kudernac2011, mishra2015, eisenhut2018, kugel2018, simpson2019, zhang2019, ren2020, wu2020}.

Although Brownian motors have been extensively studied from a theoretical point of view~\cite{astumian1997, reimann2002, hanggi2005, hanggi2009}, there are only a few examples of molecular motor modeling for systems like the ones described above.
For example, Stolz~\textit{et al.} used a statistical model with several free parameters to interpret their experiments~\cite{stolz2019}.
They assumed the probability of overcoming the potential barrier to be an error function with different characteristic energies depending on which direction the molecule is rotating.
Although this model showed a good agreement with the experimental results, it required the inclusion of several \textit{ad hoc} parameters, thus hiding the dynamical processes behind the calculated quantities.
In contrast to this, Echeverria \textit{et al.}~\cite{echeverria2014} presented a dynamical model to study molecular motors.
They assumed a noiseless overdamped dynamics of the rotational degree of freedom where the electronic state of the molecule resets periodically to its excited state with a frequency not far from that of the rotational frequency. The electronic excitation of the molecule periodically changes the potential energy surface associated with the rotational degree of freedom, which is the cause of the movement.
Although this model offers a richer view of the dynamics of molecular motors, it presents two disadvantages when modeling experiments such as the one
in Ref.~\onlinecite{tierney2011a}.
First, it assumes that the IET process directly excites the rotational degree of freedom, though in most cases a different vibrational mode first absorbs the energy from the IET event, which is then transferred to the rotational degree of freedom.
Second, according to Ref.~\onlinecite{tierney2011a}, there is a great difference in the time scales of the rotational dynamics and the waiting time between hops among the molecule's equilibrium positions.
This means that the rotation of the molecule is not a continuous process but occurs as a sequence of sudden jumps. Furthermore, the neglected role of the thermal noise and the assumption of an overdamped dynamics are not necessarily valid.

In this work, we propose a dynamical model to describe molecular motors working as Brownian machines. We assume that a fraction of the energy coming from a single IET event is responsible for the rotational excitation. As with temperature ratchets, the potential asymmetry propels the molecule in a preferential direction.
We address this behavior by explicitly solving the molecule's dynamics and performing a statistical analysis on directionality and related quantities.

Given the fact that there is an electric current flowing through the molecular device, it is fair to wonder about the role of current-induced forces (CIFs)~\cite{bode2011, fernandez2017, hoffmann2017, chatterjee2018, chen2019, lin2019, kershaw2020, mccooey2020, deghi2021, xiao2021}. These forces arise from the energy exchange that may take place between the traveling electrons and the molecular nuclei.
\footnote{The forces exerted by the electrons over the classical degrees of freedom are dubbed by some authors as current-induced forces, where the general expressions can be applied to both equilibrium and nonequilibrium situations.~\cite{bode2011} Other authors use the terms \textit{direct} and \textit{wind} forces to distinguish equilibrium and nonequilibrium electron forces, respectively.~\cite{hoffmann2017}
In the present work, we reserve the term ``CIF'' only for electron forces caused by nonequilibrium conditions between the electronic reservoirs}
In general, for rotational systems like the present one, it has been shown that CIFs can be nonconservative, and therefore contribute to the directional rotation~\cite{dundas2009, bustos2013, arrachea2015, ludovico2016, calvo2017, bustos2019, fernandez2019, ribetto2021}.
Furthermore, at the considered size scale, quantum effects may enter on both the electric currents and the CIFs, thus demanding a proper 
treatment. For this purpose, we estimate the effects of the CIFs on the molecular dynamics, by using a simple Hamiltonian model through the nonequilibrium Green's function (NEGF) formalism~\cite{bode2011, deghi2021}.

We illustrate the validity and usefulness of our model by applying it to the single-molecule electric motor of Ref.~\onlinecite{tierney2011a}, as it contains much of the relevant information necessary for the model's implementation. We show how to extract microscopic dynamical parameters of the system, and that the obtained results are consistent with the available experimental data.
We highlight that our model is general, in the sense that it could be applied to other molecular setups, if enough experimental data were available.

The manuscript is organized as follows: In Section~\ref{sec:model} we provide the general theoretical framework of our work, describing the implemented
model (Subsection~\ref{sec:sek}), the parameters used in the numerical
calculations (Subsection~\ref{sec:parameters}), details of the simulations
(Subsection~\ref{sec:simulations}), and a brief discussion about directionality (Subsection~\ref{sec:dir}). In Section~\ref{sec:simres} we show and discuss the results of the simulations. In Section~\ref{sec:cif}, we first show how CIFs can be included in the dynamical simulations (Subsections \ref{sec:cif-work} and \ref{sec:cif-model}) by means of the nonequilibrium Green's function (NEGF) formalism (Subsection~\ref{sec:cif-green}). Then, we estimate the role of CIFs in the analyzed example (Subsection~\ref{sec:cif-results}). Finally, in Section~\ref{sec:conclusions} we provide the main conclusions.

\section{\label{sec:model}Theoretical framework and simulation details}

\subsection{\label{sec:sek}Sudden energy kick model \& Langevin equation}

As mentioned before, it is possible to consider an electrical molecular motor as similar to a temperature ratchet, but where each IET event instantaneously excites a specific vibrational mode of the molecule. Next, an energy redistribution among the different vibrational modes takes this excitation to the rotational degree of freedom, giving rise to the molecular rotation.

We simplify these phenomena by introducing a model that neglects all the intermediate steps involved in the energy relaxation processes of the vibrational states. This implies that, on average, these intricate processes do not affect the subsequent rotational dynamics. Therefore, we assume that after the IET event the rotational motion is initiated by a ``sudden energy kick'' (SEK) with a fraction of the total energy coming from the tunneling event [cf. Fig.~\ref{fig:scheme}(b)]. 
In order to achieve directional motion, this kick should leave the molecule in a state
with rotational energy higher than the torsional potential amplitude $U_0$. 
As expected from the energy redistribution processes, the energy of the kick must be lower than the energy absorbed from the IET event, which we denote as $\hbar \omega_0$ (for the setup of Ref.~\onlinecite{tierney2011a} this corresponds to the C-H stretching excitation energy).
In our model, this is accounted for by defining the energy kick as $E_0^k = f \hbar \omega_0$, where $0<f<1$ represents the average fraction of the total energy that is transferred to the rotational degree of freedom.
This parameter will be determined by comparing our simulations with experimental data (see Sec. II C).

To reflect the sporadicity of the IET excitations, the time between two
consecutive kicks is assumed to be longer than the motor's relaxation time
into an equilibrium state. Consequently, each one of these events (and its
subsequent dynamics) is taken as independent from each other, allowing us to
work with ensemble averages. Assuming that the energy kick is purely kinetic,
the initial rotational speed $\dot{\theta}_0$ of the molecule can be
calculated and, together with a given initial angular position $\theta_0$, the
molecule's dynamics can be solved. This motivates us to implement a Langevin
equation approach to study the biased Brownian motion of the molecule in the
periodic and asymmetric potential landscape previously discussed [cf.
Sec.~\ref{sec:intro} and Fig.~\ref{fig:scheme}(b)]. As we will always refer to the 
rotational degree of freedom, our description of the dynamics can be obtained from an \textit{angular} or
{\textit{rotational}} Langevin equation of the form
\begin{equation}
  \mathcal{I} \, \ddot{\theta} (t) = \mathcal{F}^{\mathrm{eq}} (\theta (t)) - \gamma
  \dot{\theta} (t) + \xi (t),
  \label{eq:langevin}
\end{equation}
where $\mathcal{I}$ is the molecule's moment of inertia,
$\mathcal{F}^{\mathrm{eq}}=-\partial U^{\mathrm{eq}}/ \partial \theta$
is the deterministic torque applied to the molecule ($U^{\mathrm{eq}}$ being
the equilibrium torsional potential), $\gamma$ is the friction coefficient, and
$\xi(t)$ is a stochastic torque related to the thermal fluctuations. These
fluctuations are modeled by a stationary Gaussian noise of vanishing mean,
$\langle \xi (t) \rangle = 0$, satisfying the fluctuation-dissipation relation
$\langle \xi (t) \xi (0) \rangle = 2 A \delta (t)$ with $A =\mathcal{I} \gamma
k_B T$ the momentum-diffusion strength~\cite{hanggi2009}. In this sense, $\xi (t)$ can be expressed as
$\xi (t) = \sqrt{2\mathcal{I} \gamma k_{\mathrm{B}} T} \eta (t)$,
where $\eta (t) = \dot{W} (t)$ is a white noise and $W (t)$ a Wiener process~\cite{vanden2006}.

In order to solve Eq.~(\ref{eq:langevin}) we use the rotational equivalent of
the second-order integrator developed in Ref.~\onlinecite{vanden2006}
to determine the evolution of a system of interacting particles in the
presence of a thermal bath. For details on the rotational integrator
implemented in solving the molecule's dynamics, see
Appendix~\ref{app:integrator}.

\subsection{\label{sec:parameters}Analysis of the parameters}

The single-molecule electrical motor developed by Sykes' group~\cite{tierney2011a, murphy2014} is an ideal option to test the SEK model due to the
availability of many of the experimental parameters required for the
numerical calculations. Additionally, the CIFs analysis is considerably
simplified thanks to the simplicity of the BuSMe molecule used in the
experiment (this subject will be treated in Sec.~\ref{sec:cif}).
Complex molecules like, for example, the ones used in Ref.~\onlinecite{perera2012}, would greatly increase the difficulty of the analysis. In this case, one dimensional dynamics would not be enough~\cite{echeverria2014}, and additional calculations would be required to estimate tunneling currents.

The experiment by Stolz {\textit{et al.}}~\cite{stolz2019} could in principle be also chosen as a test case. However, the available information results in too many unknown variables for our purposes.
In particular, since the torsional potential's asymmetry is obtained within their statistical model and not from independent calculations (such as DFT simulations like in Ref.~\onlinecite{tierney2011b}), it becomes another unknown parameter for our model (besides the friction coefficient and the post-kick kinetic energy of the molecule, as we will show next).

As a first step to test our model, we gather all the relevant experimental information provided by the works of Sykes' group (see Table~\ref{tab:parameters} for a summary). The C-H stretch excitation energy has been experimentally studied and values of approximately $\hbar \omega_0 = 0.38 \ \mathrm{eV}$ have been communicated in Refs.~\onlinecite{tierney2009, tierney2011a}.
This energy value is much greater than the $0.01 \ \mathrm{eV}$ reported for the magnitude of the torsional potential amplitude of BuSMe on
Cu(111)~\cite{tierney2011a, tierney2011b}, which we denote as $U_0$.
This sawtooth-shaped potential has a period of $L = \pi / 3$ radians (or
$60^{\circ}$)~\cite{tierney2011a, tierney2011b}
and we mathematically describe it by the following piecewise function
\begin{equation}
  U^{\mathrm{eq}} (\theta) = \begin{cases}
    \dfrac{\theta}{\lambda L} U_0, & 0 \leqslant \theta < \lambda  L\\[1em]
    \dfrac{L - \theta}{L (1 - \lambda)} U_0, & \lambda L \leqslant \theta < L
  \end{cases}.
  \label{eq:sawtooth}
\end{equation}
Here, $\lambda$ is a factor that quantifies the potential's asymmetry: Its
value represents the fraction of $L$ where the potential's maximum is located, see Fig.~\ref{fig:scheme}(b).
Thus, it takes values between 0 and 1, and has to be different from $1/2$ in order to get directional motion. 
From Refs.~\onlinecite{tierney2011b, murphy2014} we estimated this parameter to be approximately $\lambda = 0.58$. Finally,
the experiments were carried out at a temperature $T = 5 \, \mathrm{K}$~\cite{tierney2011a}, and the BuSMe molecule's moment of inertia was
obtained from its geometry when adsorbed on the Cu(111) surface~\cite{tierney2011a} and estimated to be approximately $\mathcal{I}=
1.40 \times 10^{-44} \, \mathrm{kg} \cdot \mathrm{m}^{2}$.

As no information regarding the friction coefficient is provided, we will
estimate it from representative values found in the literature~\cite{croy2012, prokop2012, echeverria2014, stolz2019, lin2020, lin2021}, which lie within the range $10^{10} \lesssim \gamma \lesssim 10^{12}$
$\mathrm{s}^{-1}$. More precisely, we will use the values
$\gamma = \{10^{10}, 10^{11}, 10^{12} \} \ \mathrm{s}^{-1}$ while searching
for the best approximation to the experimental results reported in Ref.~\onlinecite{tierney2011a}.
It is possible, of course, to perform a more fine search for $\gamma$, but since our objective is to qualitatively illustrate  how the SEK model works, an estimation of the order of magnitude will be enough for our purposes.

In addition, the energy kick of the molecule, $E_0^k=f \hbar \omega_0$, is also an unknown quantity. Thus, the energy factor $f$ is another parameter (together with $\gamma$) that has to be tuned in our simulations (see Sec.~\ref{sec:simres} for the estimated $\gamma$ and $f$ values).

\begin{table}[t]
\caption{\label{tab:parameters}Parameters used in our applications of the SEK model. Unknown parameters,
such as the   friction coefficient $\gamma$ and the energy factor $f$, will be
determined after testing several combinations of their values and comparing the
simulations' data with the experimental results.}
\begin{ruledtabular}
\begin{tabular}{lcr}
    Quantity & Symbol & Value\\
    \hline
    Torsional potential amplitude\footnotemark[1]\footnotemark[2] & $U_0$ & $0.01 \ \mathrm{eV}$ \\
    \hline
    C-H stretch excitation energy\footnotemark[1]\footnotemark[3] & $\hbar \omega_0$ & $0.38 \ \mathrm{eV}$ \\
    \hline
    Torsional potential's period\footnotemark[2]\footnotemark[3] & $L$ & $\pi / 3$ (or $60^{\circ}$) \\
    \hline
    Temperature\footnotemark[1] & $T$ & $5 \ \mathrm{K}$ \\
    \hline
    Asymmetry factor\footnotemark[2]\footnotemark[4] & $\lambda$ & $0.58$ \\
    \hline
    Molecule's moment of inertia\footnotemark[1] & $\mathcal{I}$ & $1.40 \times 10^{- 44} \ \mathrm{Kg} . \mathrm{m}^{2}$ \\
    \hline
    Length of the longest arm\footnotemark[1] & $d_1$ & $559 \ \mathrm{pm}$  \\
    \hline
    Length of the shortest arm\footnotemark[1] & $d_2$ & $157 \ \mathrm{pm}$ \\
    \hline
    Angle between arms\footnotemark[1] & $\theta_{\mathrm{mol}}$ & $2.29$ (or $131^{\circ}$) \\
    \hline
    $x$ coordinate of the STM tip\footnotemark[1] & $x_\mathrm{tip}$ & $-559 \ \mathrm{pm}$ (or $-d_1$) \\
    \hline
    $y$ coordinate of the STM tip\footnotemark[1] & $y_\mathrm{tip}$ & $118.82 \ \mathrm{pm}$  \\
    \hline
    $z$ coordinate of the STM tip\footnotemark[1] & $z_\mathrm{tip}$ & $1000 \ \mathrm{pm}$   \\
    \hline
    Chemical potential difference\footnotemark[1] & $\mu_\mathrm{tip}-\mu_\mathrm{S}$ & $0.38 \ \mathrm{eV}$ \\
    \hline
    Periodicity factor\footnotemark[1] & $k$ & $6$ \\
\end{tabular}
\end{ruledtabular}
\footnotetext[1]{Taken (or estimated) from Ref.~\onlinecite{tierney2011a}.}
\footnotetext[2]{Taken (or estimated) from Ref.~\onlinecite{tierney2011b}.}
\footnotetext[3]{Taken (or estimated) from Ref.~\onlinecite{tierney2009}.}
\footnotetext[4]{Taken (or estimated) from Ref.~\onlinecite{murphy2014}.}
\end{table}

\subsection{\label{sec:simulations}Langevin dynamics simulations}

Before taking into consideration the previously gathered parameters
for the determination of the best values for $\gamma$ and $f$, we now describe the procedure used to solve this problem.
We achieve this task by following the next steps: First, we choose a specific 
value for $\gamma$ out of the three aforementioned possibilities. We
then systematically vary $f$ in the range $0 < f < 1$ and, for each value of $f$, we simulate $10^6$ SEK experiments
to calculate the system's directionality. Each one of these experiments consists in solving 
the molecule's dynamics by means of the dimensionless second-order integrator detailed in Appendix~\ref{app:integrator}. This is done by setting the initial conditions as $\theta_0 = 0$,
$|\dot{\theta}_0| = \sqrt{2 E_0^k /\mathcal{I}}$, and choosing a random
direction for the molecule's motion (i.e., CW or CCW). After that, we wait until
the molecule relaxes into one of its six possible orientations, i.e., one of
the sawtooth potential's minima. The relaxation is considered to be completed (and
the dynamics is stopped) once the molecule's total energy is lower than 10\%
of $U_0$. When this happens, the molecule's final position is registered and identified as a CW or CCW rotational event, unless the molecule does not
leave its initial position in the $\theta = 0$ well (in which case the
experiment is ignored). After all the SEK numerical experiments are performed, the hop angles distribution is constructed and the directionality is
determined. Next, the same methodology is repeated for the remaining values of $\gamma$.

With the numerical information obtained from the aforementioned procedure, we
then search for the best $f$ values, i.e., the values that provide
the closest directionality to the one reported in Ref.~\onlinecite{tierney2011a}.
Then, with every pair of $f$ and $\gamma$ values in hand, we increase the
number of SEK numerical experiments to $10^7$ with the purpose
of calculating their corresponding distribution of hop angles.
Finally, we check which one resembles qualitatively the most to the experimental
data. In this way, we estimate the order of
magnitude of $\gamma$ and finish the analysis for the unknown parameters.

\subsection{\label{sec:dir}Directionality}

In addition to provide the energy necessary for directed motion as an
external nonequilibrium source, the STM also allows to keep track of the
molecule's orientation in real time through tunneling current vs. time
($I$ vs $t$) experiments~\cite{baber2008, tierney2009, jewell2010, tierney2011a, murphy2014}. The fact that the distance
between the STM tip and the molecule's arms varies as the molecule rotates
gives rise to appreciable changes in the tunneling current. Thus, by strategically placing
the tip in the molecule's proximity and keeping it at constant height,
the authors in Ref.~\onlinecite{tierney2011a} were able to measure these changes in $I$ and
correlate them to positional changes of the molecule. This is feasible
because each of the molecule's possible orientations is linked to a
specific current range. In the special case of the BuSMe molecule, six
discrete states in the tunneling current were identified, each corresponding
to one of the six equiprobable equilibrium orientations of the molecule
with respect to the hexagonal Cu(111) surface.~\footnote{This is so in the experiment as these
orientations are \textit{inequivalent} from the point of view of the
asymmetrically placed STM tip.} Since the equilibrium positions are
equidistant, the molecule can rotate in either the clockwise or
counterclockwise direction in hops of multiples of $L= 60^{\circ}$.

In summary, obtaining several $I$ vs $t$ spectra provides a way
of measuring the direction of rotation of an individual molecule by following
its progression through its different equilibrium orientations on the surface.
With this data, the molecule's directionality can be quantified by means of
the formula~\cite{kugel2018, stolz2019}
\begin{equation}
  \mathrm{dir} = \frac{n_{\mathrm{CCW}} - n_{\mathrm{CW}}}{n_{\mathrm{CCW}} + n_{\mathrm{CW}}} 100\% ,
  \label{eq:dir}
\end{equation}
where $n_{\mathrm{CCW}}$ ($n_{\mathrm{CW}}$) is the number of counterclockwise
(clockwise) rotation events.

In Ref.~\onlinecite{tierney2011a}, the authors reported directionalities up to $- 5\%$ for one of the enantiomers
and no directional motion for the other. Moreover, by analyzing the hop angles distributions, they
found that motors with the highest directionality took more single ($\pm 60^{\circ}$) hops than double ($\pm 120^{\circ}$) or triple ($\pm 180^{\circ}$) hops, the latter being the least likely. On the other hand, motors with no directional rotation have shown almost the same probability to hop through all accessible angles. The authors also highlight that their measurements did not allow them to distinguish between $180^{\circ}$ clockwise and anticlockwise rotations, so these events were averaged out.

\begin{figure}
  \includegraphics[width=0.8\columnwidth]{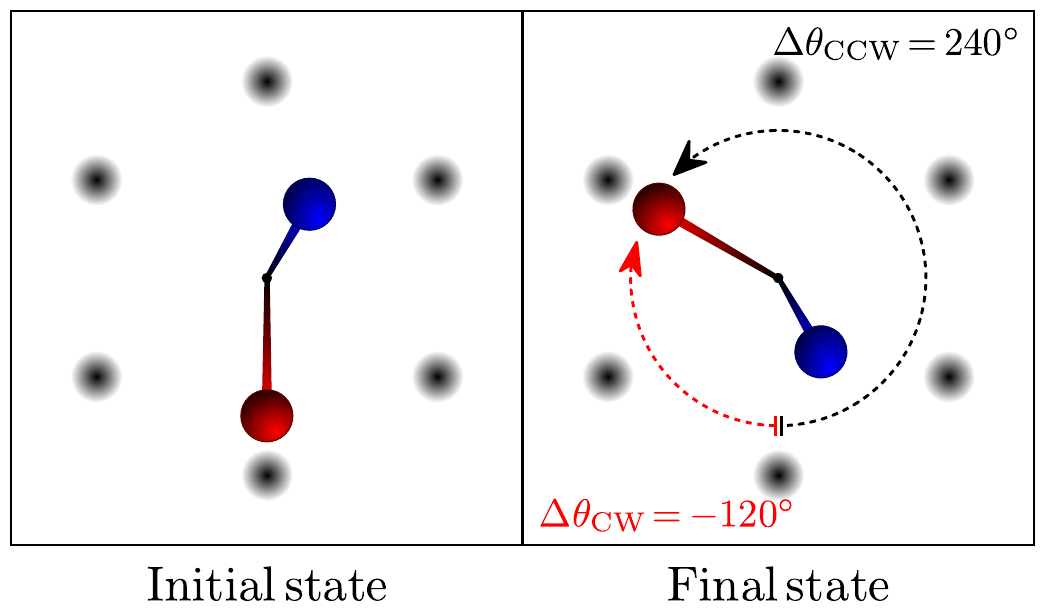}
  \caption{\label{fig:rotation} Example of two rotations, one clockwise ($\Delta \theta_\mathrm{CW}$) and one counterclockwise ($\Delta \theta_\mathrm{CCW}$), that start and end in the same equilibrium positions.
  For dynamics much faster than the observation times, rotations like these become indistinguishable, giving rise to possible rotation mislabelings.}
\end{figure}

The labeling of a rotational event as clockwise or counterclockwise is a subtle subject that deserves special attention. Since the molecule's dynamics is much faster than the time scale of the $I$ vs $t$ measurements~\cite{riedel2010, terada2010, sloan2010}, it is not possible to know with certainty in which direction the system rotates. For example, as illustrated in Fig.~\ref{fig:rotation}, a $ 240^{\circ}$ rotation could be mislabeled as a $-120^{\circ}$ rotation. For this reason, the assignment of a rotational event either as CW or CCW can lead to different results for the previous directionality definition. In particular, molecular processes such as rotations can occur on the femtosecond timescale~\cite{sloan2010} while $I$ vs $t$ experiments like the ones reported in
Refs.~\onlinecite{jewell2010, tierney2011a} have time resolutions in the millisecond scale.
Therefore, keeping track of electric current changes may be insufficient for a correct assignment of a rotation, since any change in current could be attributed to either a CW or CCW event.

Given the fact that solving the molecule's dynamics allows us to know with certainty the direction of every rotation, the above-mentioned mislabeling problem is absent in the simulations.
However, since we want to compare our results with the available experimental data, we must take into consideration this mislabeling issue.
To facilitate the discussions of the next sections, from Eq.~$\ref{eq:dir}$ we identify two directionalities:
\begin{enumerate}[leftmargin=*]
 \item The \textit{real directionality}, $\mathrm{dir}_\mathrm{r}$,
 is the one that can be obtained directly from solving the molecule's dynamics or from an ideal experiment with sufficient time resolution to avoid rotation mislabelings.

 \item The \textit{presumed directionality}, $\mathrm{dir}_\mathrm{p}$, is the directionality that would be observed experimentally considering the indistinguishability of CW and CCW rotations ending up in the same minimum, as shown in Fig.~\ref{fig:rotation}.
This directionality assumes that the molecule always takes the shortest path to its final equilibrium position.
In our simulations, we calculate $\mathrm{dir}_\mathrm{p}$ by mapping the final positions at $|\theta_f|>3L$ to the range $|\theta|\leqslant 3L$.
For example, if the molecule falls in the $\theta = 5L$ well (i.e., a $300^{\circ}$ CCW rotation) then this final position is assigned to the $\theta = -L$ well (i.e., a $60^{\circ}$ CW rotation).
For this directionality, we additionally averaged out CW and CCW $180^{\circ}$ rotations, as done in Ref.~\onlinecite{tierney2011a} for the same indistinguishability reasons. Therefore, these rotations cancel out in the numerator of Eq.~(\ref{eq:dir}), and thus their only effect is to reduce the $\mathrm{dir}_\mathrm{p}$ value. Finally, if the molecule makes complete turns (i.e., it relaxes into any of the $\theta = 6nL$ wells, where $n$ is an integer different from zero) the event is ignored since it cannot be distinguished from the case without rotation.
\end{enumerate}

Note that both directionalities will match in an ideal experiment with sufficient time resolution, or when rotations larger than $\pm 180^{\circ}$ are unlikely to happen (for example, due to energy dissipation~\cite{wu2020}).
However, unless one of these conditions is guaranteed, only the presumed directionality can be used to extract microscopic parameters via comparison between experiments and simulations. In particular, we will use this approach to estimate  the energy factor $f$.

In the following section, we illustrate our SEK model by applying it to the experimental work performed by Tierney \textit{et al.}~\cite{tierney2011a}, since it provides much of the information needed for the model's application. More specifically, by using the available experimental data together with numerical calculations, we will calculate hop angles distributions and directionalities, and show that our model is consistent with the experimental results.

\section{\label{sec:simres}Application of the SEK model}

\begin{figure*}[ht!]
  \includegraphics[width=0.9\textwidth]{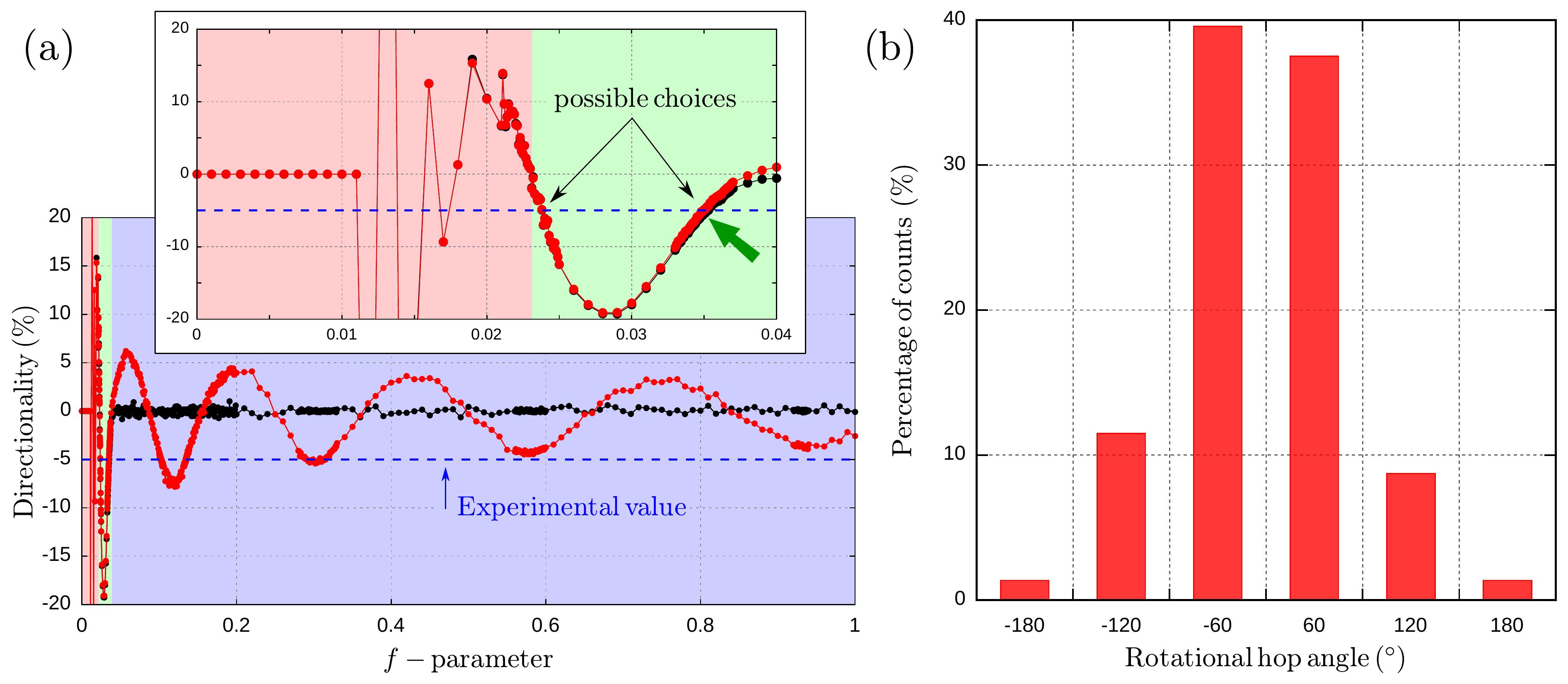}
  \caption{\label{fig:dir-f}
  (a) Real (black) and presumed (red) directionalities as functions of the energy factor $f$ for $\gamma = 10^{11} \, \mathrm{s}^{-1}$.
  The blue dashed line indicates the experimental directionality obtained in Ref.~\onlinecite{tierney2011a}. The inset is a zoom of the main figure.
  The green arrow points to the pair ($f=0.0350$, $\mathrm{dir}_\mathrm{p} = -4.87 \%$) which best reproduces the experimental data (see text for more
  details). (b) Histogram of the hop angles distribution for the $f$ value highlighted in the inset of panel (a).}
\end{figure*}

After specifying the numerical scheme designed to estimate unknown physical
parameters and discussing the system's
directionality, we now go ahead to analyze the real and presumed
directionalities for the $\gamma = 10^{11} \ \mathrm{s}^{-1}$ case.
The other values of the friction coefficient mentioned earlier, i.e.,
$\gamma = 10^{10} \ \mathrm{s}^{-1}$ and $\gamma = 10^{12} \ \mathrm{s}^{-1}$,
were discarded since they do not provide results similar to the experimental
data (see Appendix~\ref{app:dir} for details about this issue).
Fig.~\ref{fig:dir-f}(a) shows both the real and the
presumed directionalities as functions of $f$ in the $\gamma = 10^{11} \ \mathrm{s}^{-1}$ scenario.
A general pattern of three distinct behaviors associated to ranges of low,
intermediate and high values of initial energy can be appreciated.

\begin{enumerate}[leftmargin=*]
\item For the range with the smallest $f$ values [red-shaded region in
Fig.~\ref{fig:dir-f}(a) and inset] it can be seen that both directionalities are
almost identical. The reason behind this similarity is that it is very 
unlikely for the molecule to hop to wells with $\theta > 3L$ because of 
the small initial energy. The directionalities are highly
fluctuating, being strictly zero or taking random values in the whole
range. The strictly zero case can be linked most of the
times to the molecule being stuck in the $\theta = 0$ well. As $f$ increases,
there are some rare cases where the low initial energy of the molecule is
enough for it to leave its initial position. This implies that
$n_{\mathrm{CCW}}$ and $n_{\mathrm{CW}}$ are very small (much
less than $1\%$ of the total number of simulations) and, because of the
slight asymmetry of the torsional potential, their difference in
Eq.~(\ref{eq:dir}) becomes random between each set of numerical experiments,
giving rise to a marked fluctuation in the directionalities.

\item In the range of intermediate $f$ values [green-shaded region in
Fig.~\ref{fig:dir-f}(a) and inset]
the two calculated directionalities are still very similar and have a negative
bell-shaped form. The slight differences that begin to appear where this
region ends are due to the fact that final positions at $\theta > 3L$ are
more likely to happen since now the molecule has enough energy to perform these jumps. The reference
$\mathrm{dir}_{\mathrm{p}} = - 5\%$ value is intersected twice by the presumed
directionality curve, thus we get two values of $f$ for their associated hop
angles distributions to be analyzed. After a finer exploration around these
intersections, we found that the best values are $f=0.0238$ and
$f=0.0350$.

\item Lastly, we have the range with the highest $f$ values [blue-shaded region in Fig.~\ref{fig:dir-f}(a)], where $\mathrm{dir}_\mathrm{p}$ adopts an oscillatory behavior while $\mathrm{dir}_\mathrm{r} \approx 0$. On one hand, the fact that the real directionality becomes zero is expected, since now the molecule starts with a very high initial energy, such that the effects of the torsional potential become negligible and the directional preference due to the potential's asymmetry is lost.
In consequence, the number of CW and CCW become almost equal: $n_{\mathrm{CW}} \simeq n_{\mathrm{CCW}}$, and then $\mathrm{dir}_{\mathrm{r}} \approx 0$. On the other hand, the oscillations of $\mathrm{dir}_\mathrm{p}$ are just a numerical \textit{artifact} due to the fact that some CW rotations are registered as CCW (and vice versa) when the molecule has enough initial energy to reach potential wells at $| \theta | > 3L$. This effect is clearly important because $\mathrm{dir}_\mathrm{p}$ is the only directionality accessible experimentally and can lead to misinterpretations of the results.
Note that, within this range of energy, there are several $f$ values with $\mathrm{dir}_\mathrm{p}$ very close to $-5\%$, thus increasing the number of hop angles distributions to analyze.
\end{enumerate}

Many of the previously found $f$ values can be discarded from an energy distribution point of view.
If some energy is injected into a particular normal mode of the molecule then, after some time, it will be redistributed into the rest of the normal modes. Indeed, a precise knowledge of how this energy redistribution takes place would require molecular dynamics calculations, which are beyond the scope of this work. However, we can take the long-time behavior of the system as a guide, and then use the equipartition theorem.
In this case, any particular mode (e.g., a rotational mode) will receive a
fraction $1/3N$ of the injected energy, where $N$ is the number of atoms in
the molecule. Since the BuSMe molecule has 18 atoms, then there are 54 modes
and hence approximately a fraction $1/54 \simeq 0.02$ of the injected energy
goes to an individual rotational mode.
This simplified picture allows us to estimate an order of magnitude for the energy fraction $f$ and thus disregard values much larger than $f=0.02$, since they imply a huge energy transfer to a very specific normal mode, without any particular reason. 
Let us recall that the initial energy is deposited in a C-H mode, which bears no resemblance with the rotational mode and possesses a much higher frequency. Note that the order of magnitude of the estimated $f$ value places us in the intermediate region of this energy factor, where the two directionalities are very similar. This means that the mislabeling of rotational events present in $\mathrm{dir}_\mathrm{p}$ is relatively low here. Hence, we are left with only two values for the energy fraction: $f=0.0238$ and $f=0.0350$, i.e., the ones found in the intermediate region. Now, we are in position to check which one provides the hop angles distribution that best resembles the experimental results.

The histograms corresponding to the hop angles distribution for these $f$
values were plotted after performing $10^7$ SEK experiments and registering
the molecule's position after relaxation. In principle, as with the
directionality, we have two types of distributions, real and presumed,
depending on whether or not we take the previously discussed experimental
limitations into account. Nevertheless, as stated above, the comparisons
with the experimental data are made with the presumed directionality.
Besides, since both directionalities are very similar in this range of $E_0^k$, we also expect the two distributions to be similar as well.

With respect to the value $f=0.0238$, its associated presumed directionality
was found to be $\mathrm{dir}_{\mathrm{p}}=-5.29 \%$. However, in up to
$98 \%$ of the SEK experiments, the molecule was found to be stuck in the
initial $\theta = 0$ well. In the remaining experiments the molecule was almost
always found in the nearest neighbor wells, i.e., those at $\theta = \pm L$, and
very rarely in the rest. For this reason, this value of $f$ was
discarded since it is incompatible with the experimental data. On the other
hand, for $f=0.0350$, a presumed directionality of
$\mathrm{dir}_{\mathrm{p}}=-4.87\%$ was obtained, with approximately $8 \%$
of the SEK experiments ending in the $\theta = 0$ well. This value provided
the most resembling data to the experiment, and its associated histogram
corresponding to the presumed hop angles distribution is shown in
Fig.~\ref{fig:dir-f}(b). In this figure it can be seen that, as expected from
the physical experiments, there is a predominance of single hops, followed
by double and triple hops, where the latter are the scarcest. Hops greater
than $\pm 180^{\circ}$ only made up approximately $0.3\%$ of the total number
of SEK experiments, so the presumed distribution of hop angles is practically
indistinguishable from the real one. We stress that the above comparisons between
numerical and physical data are qualitative, and can be made as precise
as desired if enough experimental information is provided.

In summary, with the information obtained from the previous procedure, the
SEK model has been illustrated and the assessment of unknown parameters is
concluded. We estimate that the friction coefficient should be of the order
of $\gamma \sim 10^{11} \, \mathrm{s}^{-1}$ and that the initial energy for the rotation
should be approximately $E_0^k \approx 0.035 \hbar \omega_0$.
We highlight that this value of the friction coefficient is far from justifying an overdamped rotor's dynamics, unlike the common assumption in Brownian motors' modeling~\cite{reimann2002, hanggi2009, echeverria2014}.
Additionally, our simulations showed that the inclusion of thermal noise highly affects the motor's directionality, even in the low temperature situation considered here.

In the following section we incorporate and analyze the role of the current induced force when solving the molecule's dynamics, by following similar numerical strategies to those used so far.

\section{\label{sec:cif}Role of current-induced forces}

\subsection{\label{sec:cif-work}Work originated from CIFs}

As mentioned before, the presence of an electric current flowing through the molecule motivates us to wonder whether CIFs should be included in our description of the experiment, in addition to the already treated stochastic dynamics. Thus, in this section we analyze the role of these forces on the previously described and studied system: A rotational molecular motor treated as a Brownian temperature ratchet. Before continuing, it is important to highlight that our goal here is not the precise evaluation of CIFs of the studied system, which would require far more complex calculations based on combining density functional theory (DFT) with NEGF methods or similar techniques~\cite{lu2015, lu2019}. The objective is to gain some insight into the potential role of CIFs by means of a model that provides a rough but reasonable estimation of them.

\begin{figure}[ht!]
  \includegraphics[width=0.9\columnwidth]{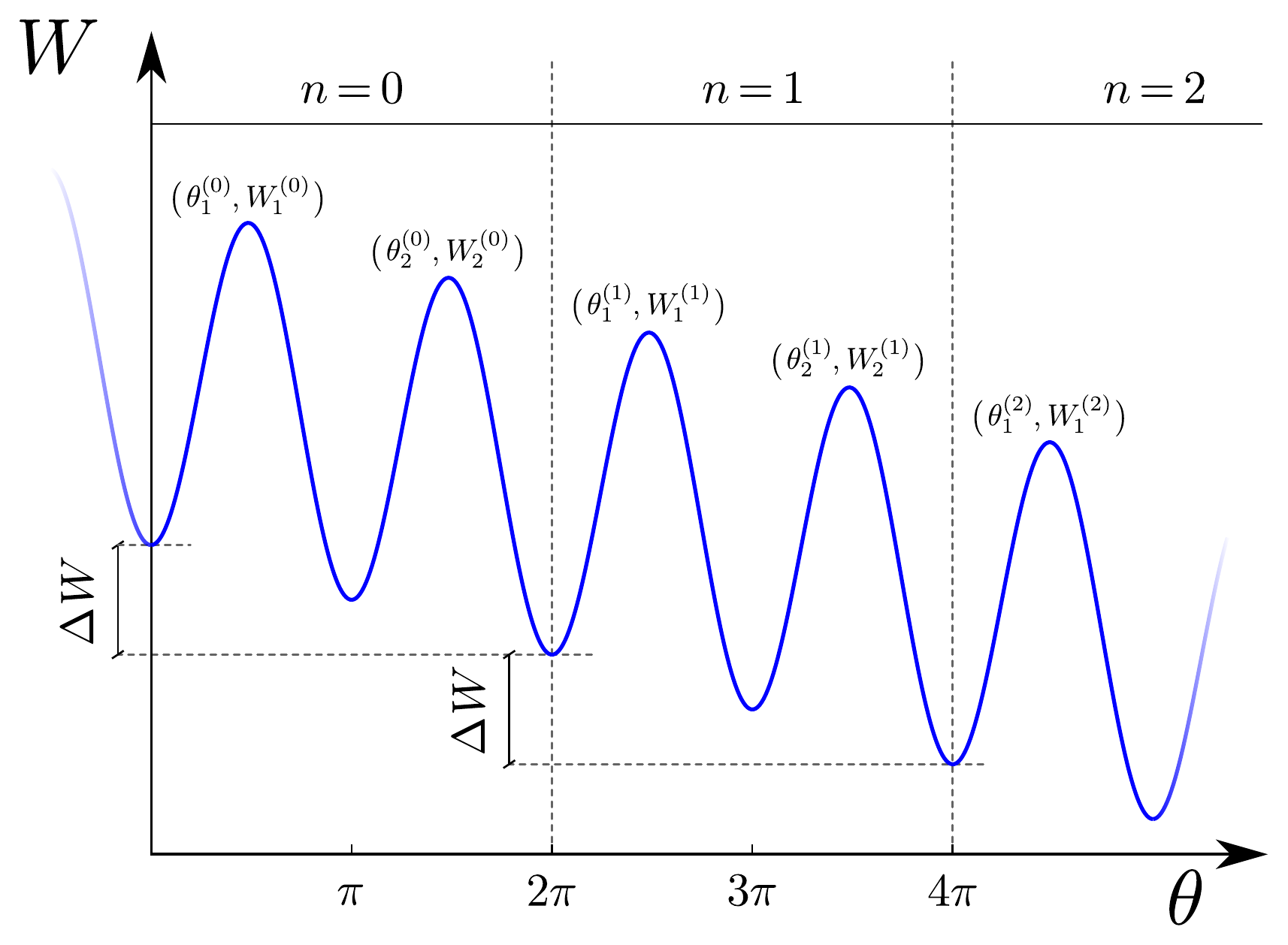}
  \caption{\label{fig:Wprofile}Example of the energy profile for a rotational motor under the
  influence of equilibrium and nonequilibrium forces. After $n$ cycles there is an energy shift of $n \Delta W$ and each
  maximum or minimum is displaced in a quantity $2\pi n$.}
\end{figure}

To begin with, we observe that the total forces or torques acting in the system can be split into two parts (cf. Sec.~\ref{sec:cif-green}): an equilibrium (or zero voltage) contribution, and a nonequilibrium one. In our model, the equilibrium contribution has already been taken as that coming from the equilibrium torsional potential, i.e., $\mathcal{F}^\mathrm{eq}(\theta)$ in Eq.~(\ref{eq:langevin}). On the other hand, the angular projection of the
CIF, denoted as $\mathcal{F}^{\mathrm{neq}}(\theta)$
(the current-induced torque), has to be added to Eq. (\ref{eq:langevin}) as a not necessarily conservative term. This results in the equation
\begin{equation}
  \mathcal{I} \ddot{\theta} (t) = \mathcal{F}^{\mathrm{eq}} (\theta(t)) + \mathcal{F}^{\mathrm{neq}}(\theta(t)) - \gamma
  \dot{\theta} (t) + \xi (t).
  \label{eq:langevin2}
\end{equation}
A direct consequence of this incorporation is the distortion of the potential energy landscape. Strictly speaking, due to the possibility of nonconservative forces acting on the system, we should no longer talk about a potential energy function but rather a work function, which can be determined by
\begin{equation}
  W(\theta) = \int_{0}^{2 \pi} \left[ \mathcal{F}^{\mathrm{eq}}(\theta) + \mathcal{F}^{\mathrm{neq}}(\theta) \right] \mathrm{d} \theta.
  \label{eq:work}
\end{equation}
In order to maintain a potential-like picture for the temperature ratchet [as the one depicted in Fig.~\ref{fig:scheme}(b)], we will be actually working with $-W(\theta)$.~\footnote{In fact, if we turn off the CIF term, then we recover the original torsional potential, i.e., $-W=U^\mathrm{eq}$.} The changes in the energy curve induced by the CIF imply that the position and amplitude of the work function's maxima and minima can be modified. Thus, these extrema must be located and calculated before solving the molecule's dynamics to assess whether the molecule is stuck in an energy well or not.
With the aim of solving this issue, we exploit the $2\pi$-cyclic behavior of $W(\theta)$, which can be written as 
\begin{equation}
W(\theta+2n\pi) = W(\theta) + n \Delta W,
\end{equation}
where $n$ is an integer number and $\Delta W = \int_{0}^{2 \pi} \mathcal{F}^{\mathrm{neq}} \mathrm{d} \theta$ represents the energy shift after one cycle (see Fig.~\ref{fig:Wprofile} for a representative example). This form obviously simplifies the simulations, as we only need to calculate $W(\theta)$ in the range $0 \leqslant \theta < 2\pi$. With this in mind, we can apply the same methodology used in Sec.~\ref{sec:simulations} for the directionality calculations.

\subsection{\label{sec:cif-model}Model for CIFs calculations}

\begin{figure}[ht!]
  \includegraphics[width=0.8\columnwidth]{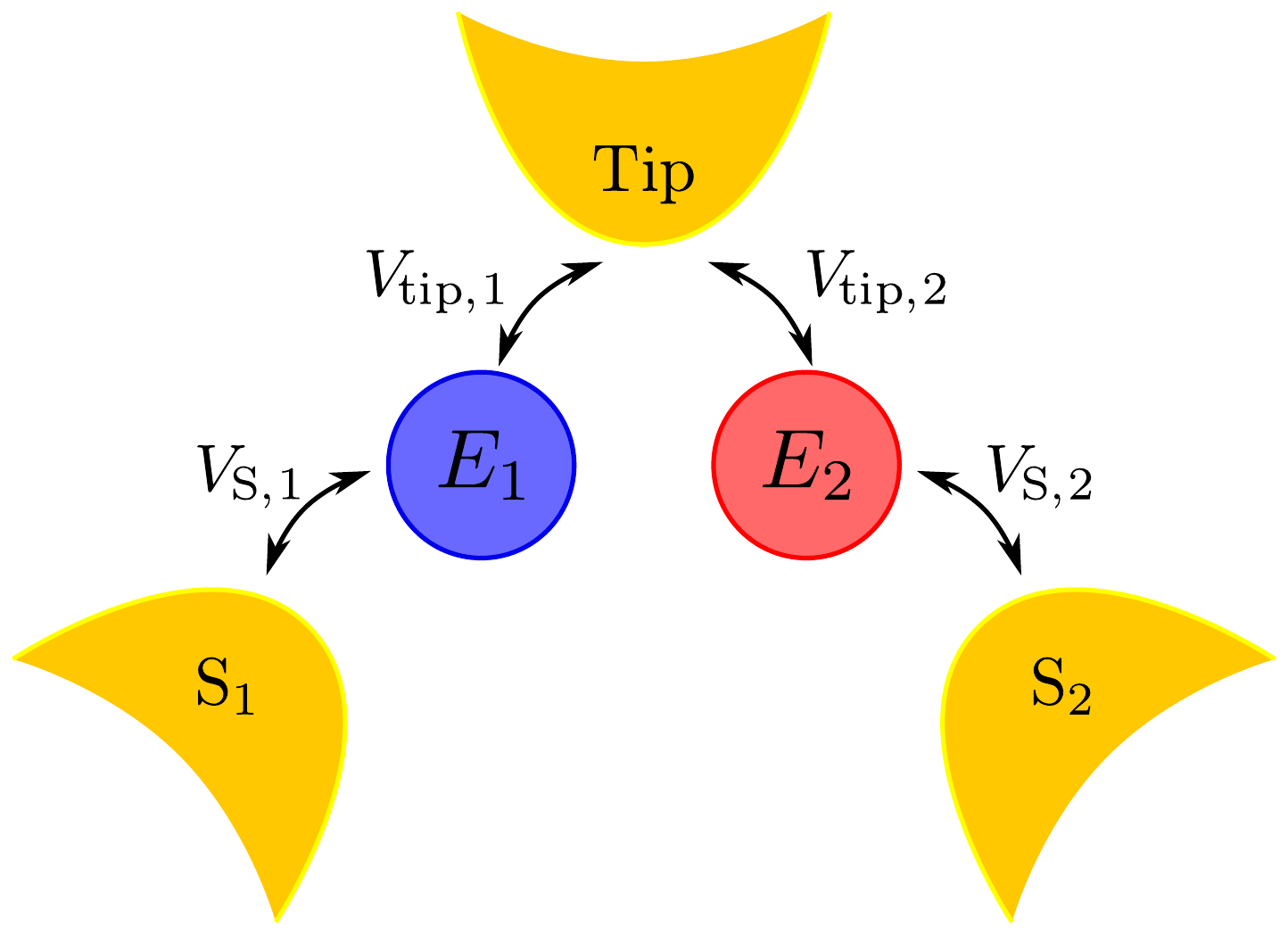}
  \caption{\label{fig:CIFscheme} Scheme of the proposed tight-binding model, consisting of two decoupled sites attached to three conduction channels: ``$\mathrm{tip}$'',  ``$\mathrm{S}_1$'' and ``$\mathrm{S}_2$''. See also Fig.~\ref{fig:scheme}(a) for geometrical details about the relative position of the sites and the STM tip.}
\end{figure}

After making clear the idea behind the calculation of $W(\theta)$ it is now
necessary to develop a microscopic model for the CIFs' calculation.
We propose a minimal tight-binding model consisting of a two sites system attached to three conduction channels through which an electrical current can flow.
Each site represents a molecular orbital through which current can flow between the tip and the metallic surface.
These orbitals are assumed to be located at the end of each of the molecule's arms, which are separated by an angle $\theta_{\mathrm{mol}}$.
The conduction channels are modeled as semi-infinite tight-binding chains within the wideband limit.
A scheme of this model can be seen in Fig.~\ref{fig:CIFscheme}, where the blue and red circles represent the system's sites, and further details can be appreciated in Fig.~\ref{fig:scheme}(a).

Sites 1 and 2 are associated with the long and short arms, respectively.
The first site is located at a distance $d_1$ from the rotation axis and at a distance $r_1$ from the STM tip.
The second one is placed at a distance $d_2$ from the rotation axis and we denote its distance to the tip as $r_2$.
The site's energies are assumed to be of the form
\begin{eqnarray}
  E_{1} & = & \bar{E} + \delta \bar{E} \cos [k (\theta + \phi)] \\
  E_{2} & = & \bar{E} + \delta \bar{E} \cos (k \theta)
\end{eqnarray}
where $\bar{E}$ is a reference energy and $\delta \bar{E}$ is the amplitude
factor related to the energy change due to the molecule's rotation with
respect to the tip's position. Since both sites' energies do not need to
be the same at every position, we add the phase difference $\phi$ between them. Finally, $k = 2\pi/L = 6$ is the periodicity factor that is linked to the torsional potential.

These tight-binding sites are coupled to a single orbital at the edge of a conduction channel labeled ``tip'', that represents the STM located at some position $\bm{r}_\mathrm{tip} =
(x_{\mathrm{tip}},y_{\mathrm{tip}},z_{\mathrm{tip}})$ and with chemical potential $\mu_\mathrm{tip}$.
Considering that the STM tip is asymmetrically placed over the molecule, each site has a distinct coupling to the tip, which we dub $V_{\mathrm{tip},1}$ and $V_{\mathrm{tip},2}$.
These couplings depend on the distance to the tip, and we model their amplitudes by the relation
\begin{equation}
  V_{\mathrm{tip},i} = t_\mathrm{tip} \exp{[a(1-r_i/r_0)]},
\end{equation}
where $i=\lbrace 1,2 \rbrace$. Here, $t_{\mathrm{tip}}$ is a characteristic
tunneling (hopping) amplitude corresponding to the case where the STM tip is closest to the site 1 (this distance being $r_0$), $a$ is a decay constant, and $r_i$ is the distance between the STM tip and the site $i$. 

Additionally, the sites are attached to two conduction channels, $\mathrm{S}_1$ and $\mathrm{S}_2$,
both with chemical potential $\mu_{\mathrm{S}}$ and representing the substrate over
which the molecule lies. For this case, we consider a constant and identical
coupling to both sites and denote it as
\begin{equation}
  V_{\mathrm{S},i} = t_\mathrm{S},
\end{equation}
again with $i=\lbrace 1,2 \rbrace$. Additional details related to this model
can be found in Appendix~\ref{sec:app-tb}.

Given the fact that the metallic surface is modeled by two independent conduction channels, electrons flowing to the substrate through different sites do not interfere.
In this way, the present tight-binding model circumvents peculiar effects such as antiresonances~\cite{damato1989,solomon2008}, which are not expected in this kind of experiments.

Several of the new parameters that appear in this section were estimated from
some of the works already discussed in Sec.~\ref{sec:parameters} and are
also displayed in Table~\ref{tab:parameters}. The model has, nonetheless,
six unknown parameters: $\bar{E}$, $\delta \bar{E}$,
$\phi$, $t_\mathrm{tip}$, $a$, and $t_{\mathrm{S}}$. They will be estimated in
Sec.~\ref{sec:cif-results} after calculating the electrical current
flowing through the system as a function of the angular variable $\theta$
and comparing the obtained results with the experimental data. Before achieving this task, it is necessary to lay the ground for the
theoretical framework required for the CIFs and electrical currents calculations. This
is the objective of the following section.

\subsection{\label{sec:cif-green}Nonequilibrium Green's functions formalism}

With the previous model in mind, we will calculate the CIFs by means of the
NEGF method. In this formalism, the mean value of the adiabatic (zeroth order) force is given by~\cite{bode2011, deghi2021}
\begin{equation}
	F_{\nu} = \int\frac{\mathrm{d} \epsilon}{2\pi i} \mathrm{Tr} [ \hat{\Lambda}_{\nu}\hat{G}^{<} ],
	\label{eq:fnu}
\end{equation}
where $\hat{\Lambda}_{\nu} \equiv -\partial\hat{H}_\mathrm{el}/\partial X_\nu$ is the force operator related to the tight-binding Hamiltonian $\hat{H}_\mathrm{el}$ and the mechanical variable $X_\nu$, and $\hat{G}^{<}$ is the lesser Green's function operator representing the electronic density. Within the Keldysh formalism, $\hat{G}^{<}$ is given by the relation~\cite{haug2008}
\begin{equation}
	\hat{G}^{<} = \hat{G}^r \, \hat{\Sigma}^{<} \, \hat{G}^a.
\end{equation}
Here, $\hat{G}^r$ ($\hat{G}^a$) is the retarded (advanced) Green's function operator and $\hat{\Sigma}^{<}$ is the lesser self-energy
\begin{equation}
	\hat{\Sigma}^{<} = 2 i \sum_\alpha f_\alpha \hat{\Gamma}_\alpha,
\end{equation}
where $\alpha = \{\mathrm{tip},\mathrm{S}_1,\mathrm{S}_2\}$ is a reservoir index, and $f_\alpha(\epsilon,\mu_\alpha,T_\alpha)$ is the
Fermi-Dirac distribution function of reservoir $\alpha$, assumed to be in equilibrium at a chemical potential $\mu_\alpha$ and temperature $T_\alpha$. Lastly, $\hat{\Gamma}_\alpha \equiv -\mathrm{Im}(\hat{\Sigma}^r_\alpha)$, where
$\hat{\Sigma}^r_\alpha$ is the retarded self-energy operator of reservoir $\alpha$.

We will think of the $f_{\alpha}$ distributions as displaced from an
equilibrium distribution $f_0$, namely $f_\alpha = f_0 + \Delta f_\alpha$. Thus, the adiabatic force in Eq.~(\ref{eq:fnu}) can be linearly decomposed in two terms: 
\begin{enumerate}[leftmargin=*]
\item An equilibrium force, which for our purposes yields the first term of Eq.~(\ref{eq:langevin}) linked to the sawtooth-shaped potential of Eq.~(\ref{eq:sawtooth}), i.e., $F_\nu^\mathrm{eq}=-\partial U^\mathrm{eq}/\partial X_\nu$. For the sake of simplicity in our analysis, and without any loss of generality, we assume that all equilibrium effects coming from the electronic Hamiltonian $\hat{H}_\mathrm{el}$ are already embedded in this force.
\item A current-induced force, defined within the NEGF formalism as~\cite{bode2011, deghi2021}
\begin{equation}
  F_\nu^\mathrm{neq} = -\frac{1}{\pi} \sum_\alpha \int_{-\infty}^\infty \mathrm{Tr} \left[ \hat{\Lambda}_\nu \hat{G}^r
  \hat{\Gamma}_\alpha \hat{G}^a \right] \Delta f_\alpha \mathrm{d} \epsilon,
\end{equation}
which takes into account the bias voltage subtended between the substrate and the STM tip.
\end{enumerate}

Obviously, we are interested in the angular equivalents of these quantities (i.e., torques), which can be easily obtained by writing $X_\nu$ in terms of $\theta$ and projecting on the angular direction $\hat{\bm{\theta}}$. Lastly, the electrical current through the conduction channel $\alpha$ can be determined within the Landauer-Büttiker
formalism~\cite{buttiker1986} by the formula
\begin{equation}
  I_\alpha = \frac{e}{h} \sum_\beta
  \int_{-\infty}^{\infty} T_{\alpha \leftarrow \beta} (f_\beta-f_\alpha) \mathrm{d}\epsilon,
  \label{eq:current}
\end{equation}
where $T_{\alpha \leftarrow \beta} = 4 \mathrm{Tr} [\hat{\Gamma}_\alpha \hat{G}^r \hat{\Gamma}_\beta \hat{G}^a]$ is the transmission coefficient between conduction channels $\alpha$ and $\beta$. As before, supplementary information about the previous expressions can be found in Appendix~\ref{sec:app-tb}. With the model described in Sec.~\ref{sec:cif-model} and the formulas presented above we now only require to determine the model's unknown parameters in order to perform the necessary calculations for the study of CIFs effects. All this will be done in the following subsection.
\begin{figure}[ht!]
  \includegraphics[width=\columnwidth]{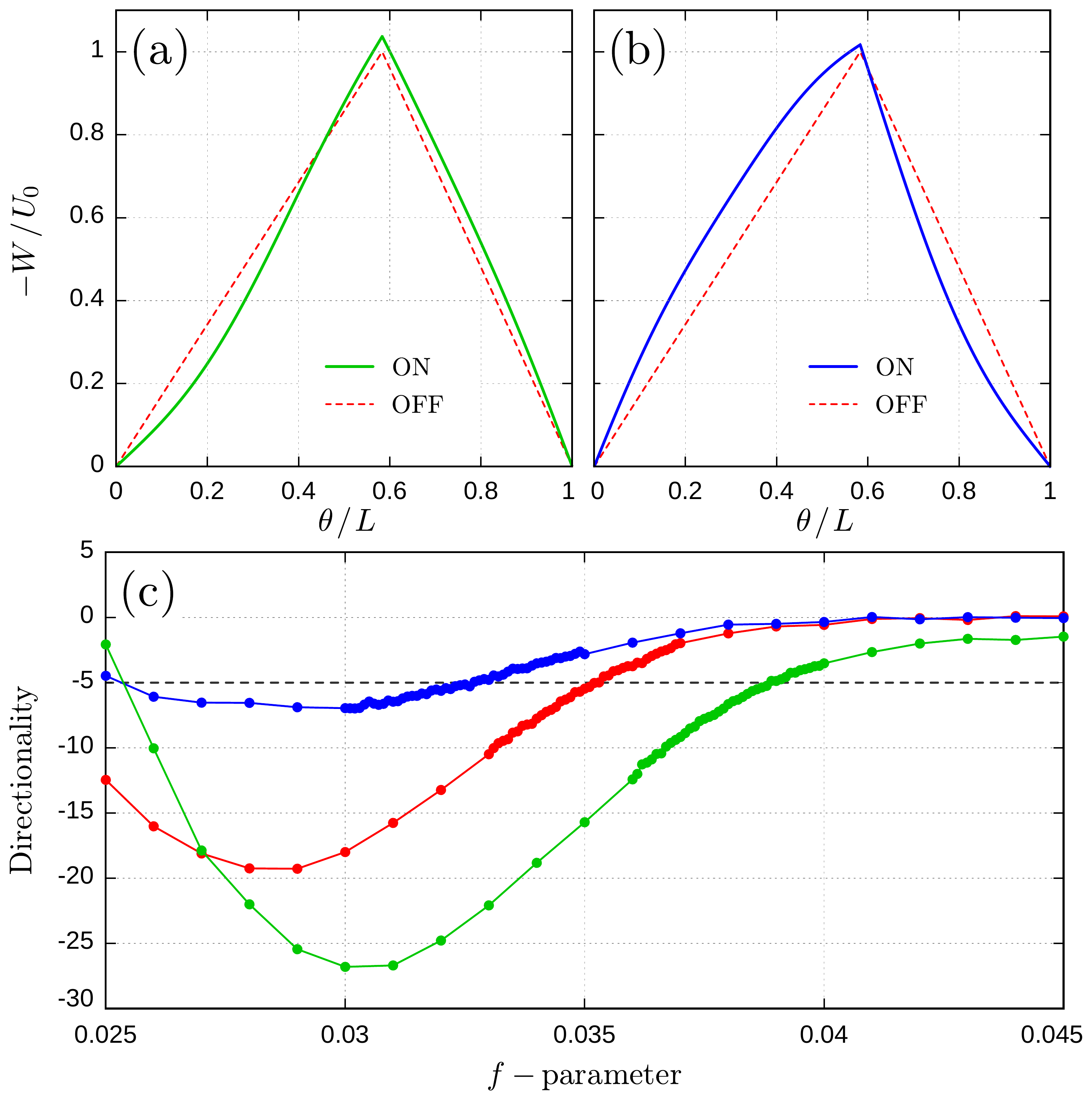}
  \caption{\label{fig:CIFlandscape}(a)
  Energy landscape for both equilibrium and nonequilibrium cases (OFF and ON, respectively).
  For the latter we used $a=2.38$, $\bar{E}=4.78 \ \mathrm{eV}$, $\delta \bar{E}=0.62 \ \mathrm{eV}$, $\phi = \pi /5$, 
  $t_\mathrm{tip}=0.02 \ \mathrm{eV}$ and $t_{\mathrm{S}} = 8.92 \ \mathrm{eV}$.
  (b) Same as (a) but with $a=2.38$, $\bar{E}=1.97 \ \mathrm{eV}$,
  $\delta \bar{E}=0.25 \ \mathrm{eV}$, $\phi = 2 \pi /5$, $t_\mathrm{tip}=0.01 \ \mathrm{eV}$ and $t_{\mathrm{S}} = 7.05 \ \mathrm{eV}$.
  (c) Presumed directionalities as functions of $f$ for: the equilibrium case (red curve)
  [i.e., same curve as shown in Fig.~\ref{fig:dir-f}(a)], and the two
  nonequilibrium scenarios mentioned before. For the case shown in (a), the
  best $f$ value is $f=0.0378$ (green curve), while for case (b) it is $f=0.0325$ (blue curve).}
\end{figure}

\begin{figure}[ht!]
  \includegraphics[width=\columnwidth]{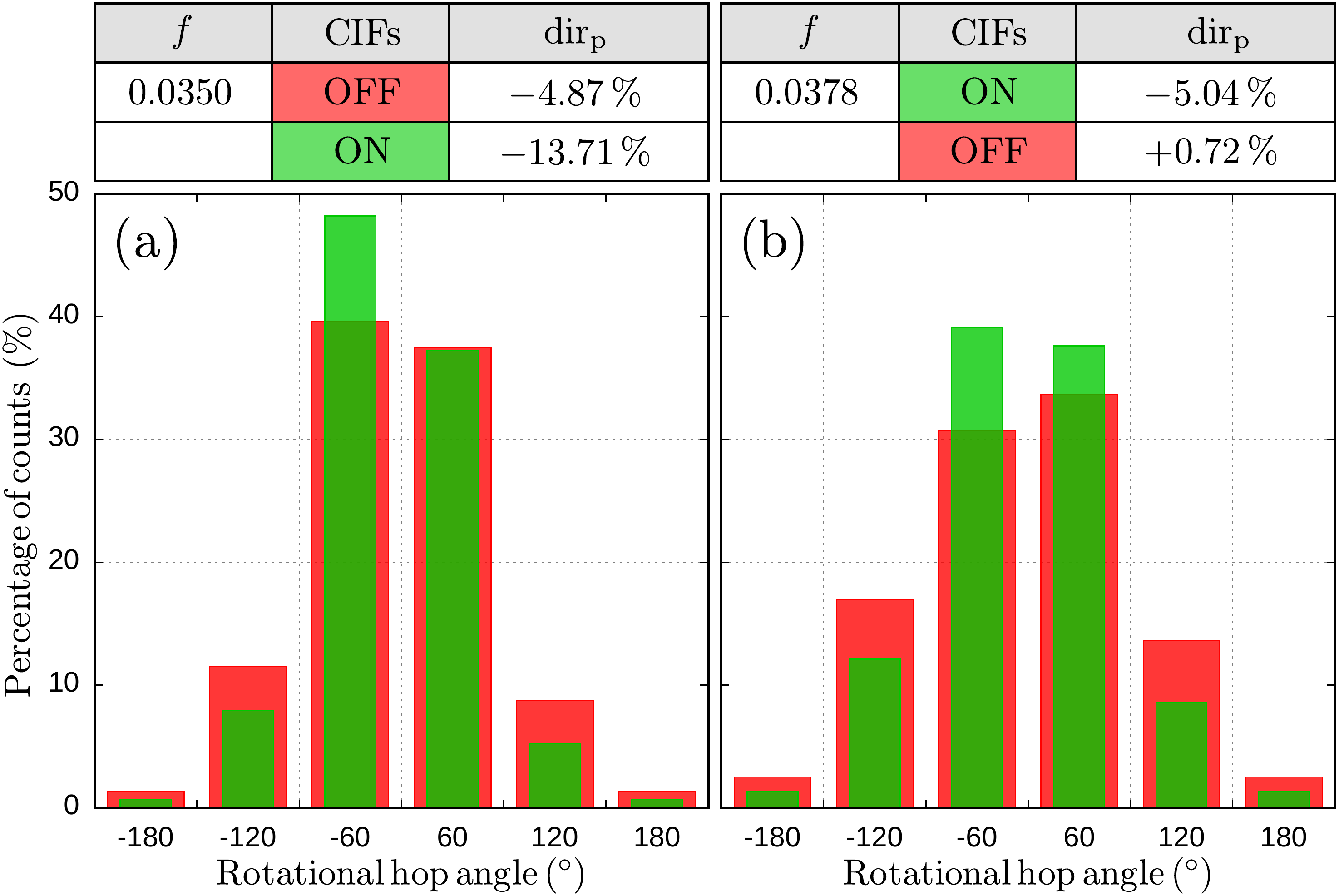}
  \caption{\label{fig:CIFhist}(a) Hop angles distributions for $f=0.0350$ with and without CIFs.
  This value of $f$ was obtained after applying the SEK model without considering CIFs
  (cf. Sec.~\ref{sec:simres}).
  (b) Hop angles distributions for $f=0.0378$ with and without CIFs. This value of
  $f$ was obtained after applying the SEK model and taking CIFs into account.
  The obtained presumed directionalities for all these cases are summarized in the tables above.}
\end{figure}

\subsection{\label{sec:cif-results}Analysis of the parameters \& Results}

As a first step for determining the model's six unknown parameters
($\bar{E}$, $\delta \bar{E}$, $\phi$, $t_\mathrm{tip}$, $a$, $t_{\mathrm{S}}$), we analyze
the experimental data regarding the tunneling current vs time measurements.
In particular, we are interested in how this current depends on the rotational
angle $\theta$.	As the authors in Ref.~\onlinecite{tierney2011a} state, the
molecule's orientations can be correlated to specific tunneling current ranges.
Thus, from the $I$ vs $t$ curve presented in that work, we estimate
the tunneling current values, $I_\mathrm{tip}^\mathrm{(exp)}$, for each of the molecule's six possible
orientations and display them in Table~\ref{tab:currents}.
\begin{table}[h]
\caption{\label{tab:currents}Experimental and numerical values for the tunneling current for the values of $\theta$ corresponding to the molecule's six possible positions. The experimental values were estimated from Ref.~\onlinecite{tierney2011a} while the numerical ones were obtained through Eq.~(\ref{eq:current}). In addition, we display the associated relative error between them.
As discussed in Refs.~\onlinecite{jewell2010, tierney2011a}, note that the highest current value is obtained when the tip is closest to the molecule (check Table~\ref{tab:parameters}).}
\begin{ruledtabular}
\begin{tabular}{cccc}
    $\theta$ ($\pi/3$) & $I_{\mathrm{tip}}^{\mathrm{(exp)}} \ ( \mathrm{pA} )$ & 
    $I_{\mathrm{tip}}^{\mathrm{(num)}} \ ( \mathrm{pA} )$ & Rel. error\\
    \hline
    $0$ & $\approx 5.2$ & $4.7$ & $9.6$\% \\
    \hline
    $1$ & $\approx 4.3$ & $4.6$ & $7.0$\% \\
    \hline
    $2$ & $\approx 9.8$ & $8.9$ & $9.2$\% \\
    \hline
    $3$ & $\approx 12.1$ & $13.1$ & $8.3$\% \\
    \hline
    $4$ & $\approx 8.0$ & $8.0$ & $0.0$\% \\
    \hline
    $5$ & $\approx 6.0$ & $5.9$ & $1.7$\% \\
\end{tabular}
\end{ruledtabular}
\end{table}

With this information now available, we started by searching the six-dimensional
space of unknown parameters for combinations that would yield current values
similar to the experimental ones. We established a tolerance of 10 \% in the
relative error between experimental and numerical values, and calculated the
tunneling current $I_\mathrm{tip}^\mathrm{(num)}$ through Eq.~(\ref{eq:current}). It is
possible, of course, to consider a smaller tolerance, which would require
a much greater numerical exploration of the parameters. However, as with the
previous comparisons between experimental and numerical data (cf. Sec.~\ref{sec:simres}), our goal here is
simply to illustrate our model.

Despite the simplicity of our tight-binding model, we were able to find many
sets of values within the proposed range of relative error. All of the
combinations found shared the fact of having no appreciable work per cycle,
i.e., $\Delta W \approx 0$. Nevertheless, the associated CIFs' distortion of
the original sawtooth potential landscape was enough to produce changes in
the corresponding hop angles distribution and, consequently, in the system's
directionality. It is worth mentioning that distortions in the energy landscape
have been observed in molecular rotors based on tetra-\textit{tert}-butyl nickel
phthalocyanine molecules on Au(111)~\cite{lu2018}, and directly linked to the presence
of the STM tip.

Figs.~\ref{fig:CIFlandscape}(a) and (b) show the energy landscape $-W(\theta)$ in a couple of
nonequilibrium cases corresponding to two of the
($\bar{E}$, $\delta \bar{E}$, $\phi$, $t_\mathrm{tip}$, $a$, $t_{\mathrm{S}}$) sets found in
the parameters exploration. For comparison, we additionally show the already
treated equilibrium case, i.e., the sawtooth potential discussed before. Both sets yield the
same tunneling current values, $I_\mathrm{tip}^\mathrm{(num)}$, for each of the six possible orientations, and
we also show them in Table~\ref{tab:currents} together with their
associated relative errors. One of the criteria for choosing these sets of
values was to keep the energy amplitude as close as possible to the torsional potential's amplitude $U_0$. This is so, because we expect that this experimentally measured quantity already includes the effects of CIFs. Finally, we also imposed the condition $t_\mathrm{S} > t_\mathrm{tip}$, since the molecule is much closer to the substrate than the STM tip.

Although the distortion of the potential seems to be small, it is important enough to
give rise to different results than those obtained without CIFs.
Fig.~\ref{fig:CIFlandscape}(c) shows the $\mathrm{dir}_\mathrm{p}^\mathrm{neq}$ vs $f$
curves associated to the nonequilibrium energy landscapes discussed
before. We add the superscript ``$\mathrm{neq}$'' to $\mathrm{dir}_\mathrm{p}$ to distinguish
nonequilibrium cases from those in equilibrium (to which we add the superscript ``$\mathrm{eq}$'').
These new directionality curves were obtained by following the procedure described in Sec.~\ref{sec:cif-work} and, due to the same arguments stated in Sec.~{\ref{sec:simres}, for $\gamma = 10^{11} \, \mathrm{s}^{-1}$. Again, for the sake of comparison, we also show the
$\mathrm{dir}_\mathrm{p}^\mathrm{eq}$ vs $f$ curve. It can be clearly appreciated that with the inclusion of CIFs, the experimentally measured directionality of $-5 \%$ is now obtained for different values of $f$. In particular, for the energy landscape displayed in Fig.~\ref{fig:CIFlandscape}(a), the best value is $f=0.0378$, which yields a presumed directionality of $\mathrm{dir}_\mathrm{p}^\mathrm{neq}=-4.98\%$,
with only $\approx 2 \%$ of the SEK experiments ending in the $\theta = 0$ well. From now on we will take this case as the nonequilibrium configuration for further comparisons with the equilibrium case.~\footnote{This choice is somewhat arbitrary, since the scenario given by the energy landscape of Fig.~\ref{fig:CIFlandscape}(b) does not change the present discussion.} 

Even though the differences with the equilibrium value of $f$ seem
tiny, the hop angles distributions and, in consequence, the presumed
directionality, are highly affected by the incorporation of CIFs. By the same
token, eliminating CIFs in the nonequilibrium case also results in appreciable
changes. In Fig.~\ref{fig:CIFhist}(a) we show how the incorporation of CIFs modifies
the equilibrium distribution [where $f=0.0350$ and $\mathrm{dir}_\mathrm{p}^\mathrm{eq}=-4.87\%$,
as already shown in Fig.~\ref{fig:dir-f}(b)], now giving rise to a different value for the presumed
directionality: $\mathrm{dir}_\mathrm{p}^\mathrm{neq}=-13.71\%$. Furthermore,
the number of dynamics ending in the $\theta = 0$ well increased to $\approx 24\%$.
Such a difference between equilibrium and nonequilibrium presumed directionalities
for the same value of $f$ shows how strong can the CIFs' effects be. Likewise, in
Fig.~\ref{fig:CIFhist}(b) we display the hop angles distribution of the nonequilibrium
case previously discussed (where $f=0.0378$ and $\mathrm{dir}_\mathrm{p}^\mathrm{neq}=-4.98\%$),
together with the histogram
corresponding to the same value of $f$ but without taking CIFs into account.
By ignoring the CIFs, now there are more CCW events than CW, and the equilibrium presumed directionality
takes the value $\mathrm{dir}_\mathrm{p}^{\mathrm{eq}}=0.72\%$. Again, there is
an increase in the number of SEK experiments ending in the $\theta = 0$ well, now
reaching $\approx 8 \%$ of the total events. Once more, the
role of these forces is evident, to the point where the directionality now
becomes positive. A Table showing the comparisons between presumed directionalities of the
previously discussed cases is shown in Fig.~\ref{fig:CIFhist}}(c).

\begin{figure}[ht!]
  \includegraphics[width=1.0\columnwidth]{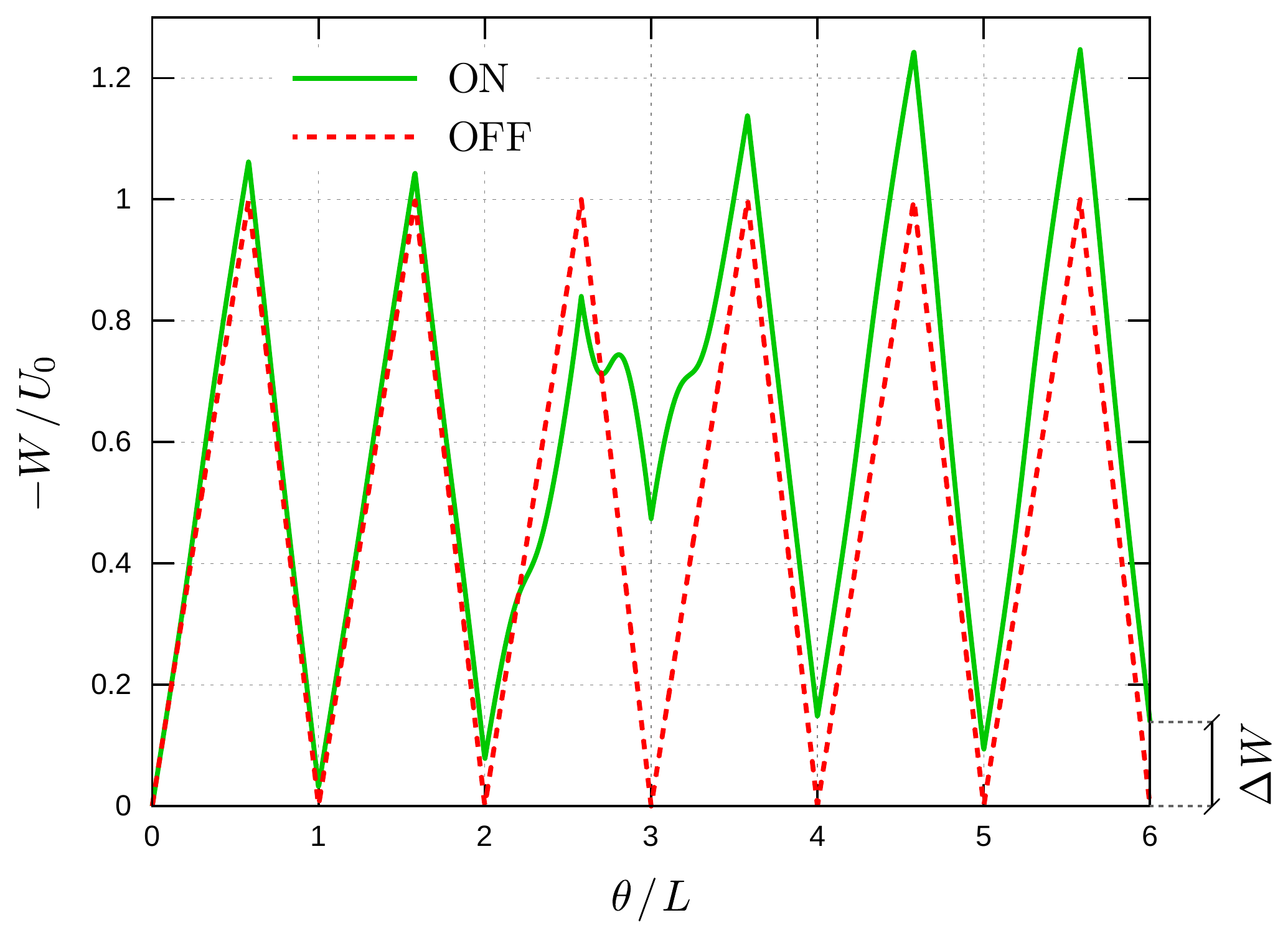}
  \caption{\label{fig:CIFW}
  Energy landscape for both equilibrium and nonequilibrium cases (OFF and ON, respectively). Here, the obtained work per cycle is approximately $\Delta W = 0.14 U_0 = 1.4$ $\mathrm{meV}$. The used parameters are: $z_\mathrm{tip} = 500$~$\mathrm{pm}$,  $t_\mathrm{tip} = 1.50$ $\mathrm{eV}$, $\bar{E} = 0.19$ $\mathrm{eV}$, $\delta \bar{E} = -0.80$ $\mathrm{eV}$, and $\phi = \pi/6$. The rest of the parameters are the same as those of Fig.~\ref{fig:CIFlandscape}.}
\end{figure}

Finally, there is an interesting aspect that the present simple model allows us to explore:
The possibility of CIFs producing useful work per cycle, in systems similar to the one exemplified here.
In particular, by neglecting the requirement of reproducing the experimental currents, it is natural to wonder about the feasibility of CIFs being nonconservative.
To address this subject, we performed a systematic exploration of the parameters' space and indeed found several sets that give rise to nonconservative CIFs. Figure~\ref{fig:CIFW} shows an example of such exploration where both  $U^\mathrm{eq}(\theta)$ and $-W(\theta)$ are plotted. In this particular case, the work per cycle is approximately $14\%$ of $U_0$, which is a considerable amount. Note also that the example's parameters are not far from those shown in Table~\ref{tab:currents} and Fig.~\ref{fig:CIFlandscape}.

Although it would be interesting to have a general recipe for predicting how each parameter affects the amount of useful work per cycle, this turned out to be a difficult nonlinear problem, despite the simplicity of the used Hamiltonian model. In this regard, this aspect deserves further study, together with a \textit{first principle} evaluation of the issue. In any case, a first observation that can be made is that strengthening the coupling between the tip and the molecule increases the chances to obtain useful work from the CIFs. As the remaining tight-binding parameters are linked to the molecule, this effect could be experimentally sought by testing different molecular compounds, or by changing the electronic occupations through the substrate’s chemical potential.

\section{\label{sec:conclusions}Conclusions}

We have proposed a simple dynamical model to study Brownian molecular motors driven by inelastic electron tunneling events. We have applied it to the first experimental proof of a single-molecule electric motor~\cite{tierney2011a}, whose results were qualitatively reproduced, proving the validity of the model and laying the ground for further explorations on similar systems. We have also demonstrated the functionality of the model by showing how it can be used to extract dynamical microscopic parameters from the experiments, thus providing a practical way of complementing the experimental information. In fact, despite the limited experimental and theoretical information available, we were able to estimate the values of some unknown parameters, such as the order of magnitude of the friction coefficient and the fraction of IET energy transferred to the rotational degree of freedom. The obtained values are reasonable, according to the bibliography and statistical arguments.

We have also shown how to incorporate CIFs into our dynamical model by using NEGF techniques. Although a precise calculation of this quantity would require DFT+NEGF calculations, we were able to estimate qualitatively the effects of CIFs on molecular machines. For this purpose, we have used a minimal Hamiltonian model whose parameters were adjusted to reproduce experimental values of electric currents for different positions of the molecule.
We obtained several reasonable sets of Hamiltonian parameters which fairly reproduced the experimental currents. In all cases, we observed that they qualitatively point in the same direction: Although CIFs do not provide a significant amount of useful work per cycle, these forces do modify the system's energy landscape. Hence, CIFs affect the molecular motor's dynamics and, with it, the distribution of rotational hop angles and directionality. In fact, we showed that CIFs can even change the direction in which the motor turns. These results highlight the importance of taking CIFs into account for describing Brownian molecular motors.

One interesting aspect that deserves further study is that of the tip's structure.~\cite{tierney2011c} This structure may alter the way in which the electrons are being tunneled to the molecule, depending on the position and orientation of the latter. This means, for example, that structured tips with certain chirality can change the effective potential, thus affecting the molecular rotation depending on how the (chiral) molecule was adsorbed on the substrate. 
In our CIF calculations, this would imply a more sophisticated modeling of the tip, such as a multi-channel lead  that allows electrons to tunnel from different positions of it.
Finally, we found some scenarios where CIFs can lead to significant contributions to the work done per cycle.
Since systems sharing this characteristic can be regarded as hybrid nanomachines: part Brownian motors and part adiabatic quantum motors~\cite{bustos2013}, this opens exciting directions for further investigation.
In this respect, the present work paves the way for studying this kind of systems.

\begin{acknowledgments}
We acknowledge financial support by Consejo Nacional de Investigaciones Cient\'ificas y T\'ecnicas (CONICET); Secretar\'ia de Ciencia y Tecnolog\'ia de la Universidad Nacional de C\'ordoba (SECYT-UNC); and Agencia Nacional de Promoción Científica y Tecnológica (ANPCyT, PICT-2018-03587).
\end{acknowledgments}

\section*{Data Availability Statement}

The data that support the findings of this study are available from the corresponding author upon reasonable request.

\appendix

\section{\label{app:integrator}Second-order integrator}

In this Appendix we show the angular equivalent of the
second-order integrator proposed in Ref.~\onlinecite{vanden2006} for
solving the Langevin equation. This is simply obtained by projecting the
original integrator on an angular direction $\hat{\bm{\theta}}$, resulting in
\begin{equation}
  \left\{ \begin{array}{lll}
    \theta_{t + \Delta t} & = & \theta_t + \dot{\theta}_t \Delta t +
    C_t , \\
    \dot{\theta}_{t + \Delta t} & = & \dot{\theta}_t + \dfrac{\Delta t}{2}
    \left[ \dfrac{\mathcal{F}_{t + \Delta t}}{\mathcal{I}} +
    \dfrac{\mathcal{F}_t}{\mathcal{I}} \right] \\ & & - \Delta t \gamma
    \dot{\theta}_t + \sigma \sqrt{\Delta t} \chi_t - \gamma C_t, 
  \end{array} \right.
  \label{integrador}
\end{equation}
with $\sigma \equiv \sqrt{2 k_B T \gamma /\mathcal{I}}$ and
\begin{eqnarray}
  C_t & \equiv & \frac{\Delta t^2}{2} \left[ \frac{\mathcal{F}_t}{\mathcal{I}} - \gamma \dot{\theta}_t \right] +
  \sigma \Delta t^{3 / 2} \left[ \frac{\chi_t}{2} + \frac{\zeta_t}{2
  \sqrt{3}} \right], 
\end{eqnarray}
where $\chi_t$ and $\zeta_t$ are two uncorrelated Gaussian random
numbers. The subindexes ``$t$'' and ``$t+\Delta t$'' below each quantity indicate
the time at which it is being evaluated.

Due to the fact that the parameters involved in this type of system
are different in many orders of magnitude (e.g., $\mathcal{I} \sim 10^{- 44}
\, \rm{kg{\cdot}m^2}$ and $\gamma \sim 10^{11} \, \rm{s}^{-1}$), from
a numerical point of view it is convenient to work with dimensionless
parameters. Taking this into consideration, we obtain the dimensionless
version of Eq.~(\ref{eq:langevin})
\begin{equation}
  \ddot{\theta}^{\ast}_{t^{\ast}} = \frac{\mathcal{F}^{\ast}_{t^{\ast}}}{\mathcal{I}^{\ast}} - \gamma^{\ast}
  \dot{\theta}^{\ast}_{t^{\ast}} + \sqrt{\frac{2 \gamma^{\ast}
  D^{\ast}}{\mathcal{I}^{\ast}}} \eta^{\ast}_{t^{\ast}},
\end{equation}
and of Eq. (\ref{integrador})
\begin{equation}
  \left\{ \begin{array}{lll}
    \theta^{\ast}_{t^{\ast}+ \Delta t^{\ast}} & = & \theta^{\ast}_{t^{\ast}}
    + \dot{\theta}^{\ast}_{t^{\ast}} \Delta t^{\ast} +
    C^{\ast}_{t^{\ast}} , \\
    \dot{\theta}^{\ast}_{t^{\ast} + \Delta t^{\ast}} & = &
    \dot{\theta}^{\ast}_{t^{\ast}} + \dfrac{\Delta
    t^{\ast}}{2\mathcal{I}^{\ast}} [\mathcal{F}^{\ast}_{t^{\ast} + \Delta t^{\ast}}
    +\mathcal{F}^{\ast}_{t^{\ast}}] \\ & &
    - \Delta t^{\ast} \gamma^{\ast} \dot{\theta}^{\ast}_{t^{\ast}}
    + \sigma^{\ast} \sqrt{\Delta t^{\ast}} \chi_t -
    \gamma^{\ast} C^{\ast}_{t^{\ast}},
  \end{array} \right.
\end{equation}
where the $\ast$ superscript indicates that the corresponding quantity has
no dimensions. With respect to the sawtooth potential described in
Eq.~(\ref{eq:sawtooth}), its dimensionless form can be expressed as
  \begin{equation}
    U^{\mathrm{eq}\ast} (\theta^{\ast})  = \begin{cases}
      \dfrac{\theta^{\ast}}{\lambda}, & 0 \leqslant \theta^{\ast} < \lambda\\[1em]
      \dfrac{1 - \theta^{\ast}}{1 - \lambda}, & \lambda \leqslant
      \theta^{\ast} < 1
    \end{cases}
    \label{eq:U^eq}
  \end{equation}
The definitions of these dimensionless quantities are given in Table~\ref{tab:adim}, and were obtained after choosing a set of {\textit{characteristic}} parameters: an angular distance $\theta_0 = L$
and an energy $U_0$ corresponding to the sawtooth potential's periodicity
and amplitude, respectively; and a time $t_0 =\mathcal{I} \gamma \theta_0 /
U_0$ corresponding to the time it takes the system to travel an angular
distance $\theta_0$ in stationary conditions (i.e. zero acceleration). Note
that the characteristic time is not only related to the sawtooth potential
but it also depends on properties of both the molecule and the molecule's
environment.

\begin{table}[h!]
\caption{\label{tab:adim}Dimensionless quantities and their definitions.}
\begin{ruledtabular}
\begin{tabular}{cc}
    Dimensionless quantity & Formula\\
    \hline
    Time & $t^{\ast} = \dfrac{U_0}{\mathcal{I} \gamma \theta_0} t$\\
    \hline
    Angular position & $\theta^{\ast} = \dfrac{\theta}{\theta_0}$\\
    \hline
    Torque & $\mathcal{F}^{\ast} = \dfrac{\theta_0}{U_0} \mathcal{F}$\\
    \hline
    Moment of inertia & $\mathcal{I}^{\ast} = \dfrac{\theta_0^2}{t_0^2 U_0}
    \mathcal{I}$\\
    \hline
    Friction coefficient & $\gamma^{\ast} = t_0 \gamma$\\
    \hline
    White noise & $\eta^{\ast} = \sqrt{t_0} \eta$\\
    \hline
    Factor $D^{\ast}$ & $D^{\ast} = \dfrac{k_B T}{U_0}$\\
    \hline
    Factor $\sigma^{\ast}$ & $\sigma^{\ast} = \dfrac{t_0^{3
    / 2}}{\theta_0} \sigma$\\
    \hline
    Factor $C^{\ast}$ & $C^{\ast} = \dfrac{1}{\theta_0} C$\\
\end{tabular}
\end{ruledtabular}
\end{table}

\section{\label{app:dir}Analysis of directionality and hop angles distributions}

\begin{figure}[h!]
  \includegraphics[width=\columnwidth]{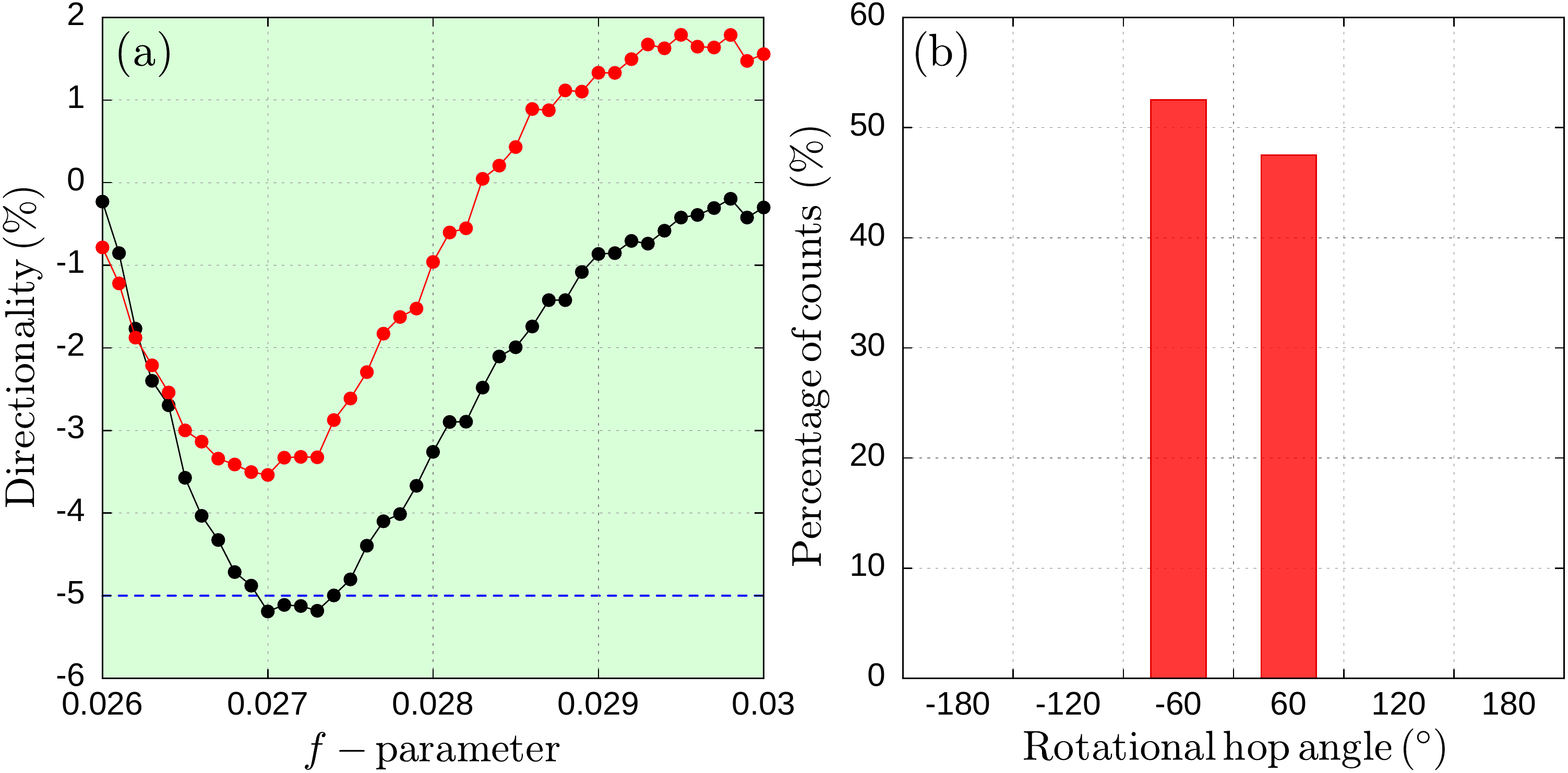}
  \caption{\label{fig:app}(a) Real (black) and presumed (red) directionalities as functions of the energy
  factor $f$ for $\gamma = 10^{10} \, \mathrm{s}^{-1}$.
  (b) Histogram showing the hop angles distribution for $\gamma = 10^{12} \, \mathrm{s}^{-1}$ and $f =
  0.129$. Since only single hops were obtained here, the real and presumed
  directionalities are equal.}
\end{figure}

The real and presumed directionalities for the $\gamma = 10^{10}$
$\mathrm{s}^{-1}$ case are displayed in Fig.~\ref{fig:app}(a). We only show
the region of intermediate $f$ values where the directionalities take the bell
shape discussed in the main text (see Sec.~\ref{sec:simres}).
In this figure it can be seen that the pressumed directionality (which is the 
one to be compared with experiments) does not intersects the experimental value 
of $-5 \%$. For this reason, we discard this value of $\gamma$.
The fact that the presumed directionality highly differs from the real one
is a consequence of the restriction that leads to the former's definition:
only considering hops smaller or equal than $180^{\circ}$. This
value of $\gamma$ provides a low friction environment which allows the molecule
to easily achieve hops greater than $180^{\circ}$ and to even make some
full turns. In consequence, both directionalities move away from each other even
in this zone of intermediate values of $f$.

Finally, we have the case of $\gamma = 10^{12} \ \mathrm{s}^{-1}$. This value
gives rise to a high friction environment for the molecule where only single
hops were obtained even for values of $f$ much higher than the ones found for
$\gamma = 10^{10} \ \mathrm{s}^{-1}$ and $\gamma = 10^{11} \ \mathrm{s}^{-1}$.
Fig.~\ref{fig:app}(b) shows the hop angles distribution for $f=0.129$, the
energy factor that provides the closest value of presumed directionality to
the experimental $-5\%$. Although it was possible to find a value of $f$ that
provides an acceptable directionality, the obtained distribution shows
that the molecule is only capable of performing simple hops. This is an
important difference from the experiment, so we discard this value of
$\gamma$.

\section{\label{sec:app-tb}Tight-binding model for a decoupled DQD attached to three conduction channels}

Here we will discuss in more detail the tight-binding model showed in
Sec.~\ref{sec:cif-model}. The Hamiltonian we considered is of the form
\begin{equation}
  \hat{H}_\mathrm{el} = \hat{H}_\mathrm{sys} + \sum_\alpha ( \hat{H}_\alpha + \hat{V}_\alpha ),
\end{equation}
where $\alpha=\{\mathrm{tip},\mathrm{S}_1,\mathrm{S}_2 \}$ is the reservoir index and 
$\hat{H}_{\mathrm{sys}}$ is the local system's Hamiltonian, including
the sites' energies. Here $\hat{H}_{\alpha}$ stands for the Hamiltonians
of the reservoirs, which are all considered to be identical and modeled as
semi-infinite tight-binding chains with site energies $E_0 = 0$ and hoppings
$V_0 \rightarrow \infty$ (wideband limit). Lastly, $\hat{V}_\alpha$
represents the coupling to channel $\alpha$.

By using the decimation technique~\cite{pastawski2001, cattena2014} the
matrix representation of the associated eigenvalue equation can be reduced
to an effective system of finite dimension:
\begin{widetext}
\begin{equation}
  [\epsilon \mathbf{I}- \mathbf{H}(\epsilon)]_{\mathrm{eff}} = \left(\begin{array}{ccccc}
    \epsilon - (E_0 + \Sigma^r(\epsilon)) & V_{\mathrm{tip},1} & V_{\mathrm{tip},2} & 0 & 0\\
    V_{\mathrm{tip},1} & \epsilon - E_{1} &  & 0 & V_{\mathrm{S},1}\\
    V_{\mathrm{tip},2} & 0 & \epsilon - E_{2} & V_{\mathrm{S},2} & 0\\
    0 & 0 & V_{\mathrm{S},2} & \epsilon - (E_0 + \Sigma^r (\epsilon)) & 0\\
    0 & V_{\mathrm{S},1} & 0 & 0 & \epsilon - (E_0 + \Sigma^r (\epsilon))
  \end{array}\right).
  \label{eq:effH}
\end{equation}
\end{widetext}

The sites connected to the conduction channels are corrected by the retarded self-energy
$\Sigma^r(\epsilon)$, and it is given by $\Sigma^r(\epsilon) = \Delta(\epsilon) - i \Gamma(\epsilon)$, with
\begin{eqnarray}
  \Delta (\epsilon) & = & \left\{ \begin{array}{ll}
    \frac{\epsilon - E_0}{2} - \sqrt{\left( \frac{\epsilon - E_0}{2}
    \right)^2 - V_0^2} &, \epsilon-E_0 \geqslant 2 V_0\\

    \frac{\epsilon - E_0}{2} &, |\epsilon-E_0| \leqslant 2 V_0 \\
    \frac{\epsilon - E_0}{2} + \sqrt{\left( \frac{\epsilon - E_0}{2} \right)^2 - V_0^2} &, \epsilon-E_0 \leqslant -2 V_0
  \end{array} \right.
\end{eqnarray}
\begin{eqnarray}
  \Gamma (\epsilon) & = & \left\{ \begin{array}{ll}
    \sqrt{V_0^2 - \left( \frac{\epsilon - E_0}{2} \right)^2} &, |\epsilon-E_0| \leqslant 2 V_0 \\
    0 &, |\epsilon-E_0| > 2 V_0
  \end{array} \right.
\end{eqnarray}
With this information, the retarded Green's function can be directly obtained
by inverting the matrix in Eq.~(\ref{eq:effH}):
$\mathbf{G}^{r}(\epsilon) = [\epsilon\mathbf{I} - \mathbf{H}(\epsilon)]_\mathrm{eff}^{-1}$. Once $\mathbf{G}^r$ is known,
the advanced Green's function can be obtained through the relation
$\mathbf{G}^a = [\mathbf{G}^r]^\dag$. Now we are able to calculate the total current flowing through the system by the expression
\begin{equation}
  I_\mathrm{tip} = \frac{e}{h} \int_{-\infty}^{\infty}
  \left( T_{\mathrm{tip},\mathrm{S}_1}+T_{\mathrm{tip},\mathrm{S}_2} \right) \left( f_\mathrm{tip}-f_\mathrm{S} \right) \mathrm{d}\epsilon ,
\end{equation}
where the transmission amplitudes are given by
\begin{equation}
  T_{\mathrm{tip},\mathrm{S}_i} = 4 \mathrm{Tr} \left[ \bm{\Gamma}_\mathrm{tip} \mathbf{G}^{r} \bm{\Gamma}_{\mathrm{S}_i} \mathbf{G}^{a} \right].
\end{equation}
Note that $I_\mathrm{tip}$ includes the currents flowing through both channels $\mathrm{S}_1$ and $\mathrm{S}_2$, the two of them having the same Fermi-Dirac distribution $f_\mathrm{S}$.

\bibliography{cite-v10p2}

\end{document}